\documentclass[prd,tightenlines,nopacs,nofootinbib,amsfonts,amssymb,amsmath,twocolumn]{revtex4}
\usepackage{graphicx}
\usepackage[usenames]{color}
\usepackage{dsfont}
\usepackage{hyperref}
\usepackage{bm}
\newcommand{\dd}{{\rm d}}

\newcommand{\be}{\begin{equation}}
\newcommand{\ee}{\end{equation}}

\newcommand{\Msol}{{~ M}_\odot}
\newcommand{\Mpc}{{~\rm  Mpc}} 
\newcommand{\hMsol}{{~h^{-1} M}_\odot}
\newcommand{\hGpc}{{~h^{-1}\rm  Gpc}} 
 
\newcommand{\hMpc}{{~h^{-1}\rm  Mpc}} 
\newcommand{\hMpcc}{{~h^{-3}\rm  Mpc^3}} 
\newcommand{\hkpc}{{~h^{-1}\rm kpc}}

\def \D {\tilde{\nabla}}

\newcommand{\<}{\langle}
\renewcommand{\>}{\rangle}

\renewcommand{\url}[1]{[#1]}

\begin{document}

\title{(Mis-)Interpreting supernovae observations in a lumpy universe}
\author{Chris Clarkson$^{1}$}
\author{George F.R. Ellis$^{1}$}
\author{Andreas Faltenbacher$^{2}$}
\author{Roy Maartens$^{2,3}$},
\author{Obinna Umeh$^{1}$}
\author{Jean-Philippe Uzan$^{4,1,5}$}
\affiliation{
$^1$Astrophysics, Cosmology \& Gravity Centre, and, Department of Mathematics \& Applied Mathematics, University of Cape Town, Cape Town 7701, South Africa\\
$^2$Physics Department, University of the Western Cape,
Cape Town 7535, South Africa \\
$^3$Institute of Cosmology \& Gravitation, University of Portsmouth, Portsmouth PO1 3FX, United Kingdom\\
$^4$Institut d'Astrophysique de Paris, UMR-7095 du CNRS,
Universit\'e Pierre et Marie Curie, 75014 Paris, France\\
$^5$National Institute for Theoretical Physics,
Stellenbosch 7600, South Africa}

\begin{abstract}
Light from `point sources' such as supernovae is observed with a beam width of order of the sources' size~-- typically less than 1 AU. Such a beam probes matter and curvature distributions that are very different from coarse-grained representations in N-body simulations or perturbation theory, which are smoothed on scales much larger than 1~{AU}. The beam typically travels through unclustered dark matter and hydrogen with a mean density much less than the cosmic mean, and through dark matter halos and hydrogen clouds. Using N-body simulations, as well as a Press-Schechter approach, we quantify the density probability distribution as a function of  beam width and show that, even for Gpc-length beams of 500\,kpc diameter, most lines of sight are significantly under-dense. From this we argue that modelling the probability distribution for AU-diameter beams is absolutely critical. 
Standard analyses predict a huge variance for such tiny beam sizes, and nonlinear corrections appear to be non-trivial. It is not even clear whether under-dense regions lead to dimming or brightening of sources, owing to the uncertainty in modelling the expansion rate which we show is the dominant contribution. By considering different reasonable approximations which yield very different cosmologies we argue that modelling ultra-narrow beams accurately remains a critical problem for precision cosmology. This could appear as a discordance between angular diameter and luminosity distances when comparing SN observations to BAO or CMB distances. 
\end{abstract}

 \date{\today}
 \maketitle

\section{Introduction: on narrow beams}

Supernovae Ia (SNIa) observations play a critical role in the evidence
for  a nonzero cosmological  constant~(see \cite{SNIa}  and references
therein).  SNIa  are  effective   standard  candles  (we  think  their
intrinsic luminosity can be calibrated from their light curve). In the
standard  cosmological model,  their  observed luminosity  is used  to
infer  their  luminosity   distance  (or  equivalently  magnitude)  by
assuming  that the  geometry of  the universe  is well-described  by a
Friedmann-Lema\^{\i}tre  (FL) background  geometry that  describes the
universe  smoothed  on  large  scales.   Of  course,  photons  do  not
propagate in  the geometry of a  smooth FL spacetime but  in the lumpy
universe:  a  beam mostly  propagates  in  underdense regions  between
clustered  matter  (overdense  islands  of matter).   The  problem  of
quantifying the  effects of  propagation in an  inhomogeneous universe
was  first  addressed,  as  far  as we  are  aware,  independently  by
Zel'dovich \cite{zel} and  Feynman \cite{feyn}.  Zel'dovich introduced
the empty-beam  approximation to deal  with light rays  propagating in
vacuum and this was extended to the case of a partially-filled beam by
\cite{dash}. These results later came  to be known as the Dyer--Roeder
approximation~--  see below.  Zel'dovich's insight  also led  to other
work on the problem in the 1960s~\cite{Bertotti,gunn,kant,refs}.

Light  propagation  in inhomogeneous  spacetimes  gives  rise both  to
distortion  of images  and magnification  of some  images,  because of
gravitational    lensing;   in    compensation,   most    images   are
demagnified. These effects induce,  in particular, a dispersion of the
observed SNIa  luminosities and hence  an extra scatter in  the Hubble
diagram~\cite{kantowski, jf97, Wang:1998eh, Wang:1999bz, Wang:2004ax}.
``Precision   cosmology"  within  the   standard  approach   could  be
compromised by  the effects of  lensing on the interpretation  of SNIa
data -- and thus it is  crucial to characterise the magnitude of these
effects  precisely. A  related  key question  is  ``what physical  and
angular  sizes  are  relevant  in  estimating these  effects  on  SNIa
observations?''

A perturbative approach (i.e. with light propagating in a perturbed FL
spacetime)  shows that  the  dispersion due  to large-scale  structure
becomes  comparable to  the intrinsic  dispersion for  redshifts  $z >
1$~\cite{hl}. However,  since the matter  fluctuations responsible for
the magnification of the SNIa also  induce a shearing of the images of
background    galaxies,    this    dispersion    can    actually    be
corrected~\cite{Cooray:2005yp,  Dodelson:2005zt}. A  similar  idea was
pursued     in      the     context     of      gravitational     wave
sirens~\cite{Shapiro:2009sr}. Nevertheless, a considerable fraction of
the lensing dispersion arises from sub-arcminute scales, which are not
probed by shear maps smoothed on arcminute scales~\cite{Dalal2002}.  

To estimate the dispersion induced by inhomogeneities, one first needs
to determine the  typical size of the geodesic  bundle associated with
SNIa. The  typical observational aperture  is of order  $1''$, whereas
the beam  is actually much thinner:  ${\cal O}(1)$ AU for  a source at
redshift $z\sim1$, i.e. an  aperture of $\beta\sim 10^{-7}{}''$.  This
is  typically  smaller than  the  mean  distance  between any  massive
objects (galaxies,  stars, H clouds,  small dark matter  halos) {--
  and on a  scale where the fluid continuum model  may not be suitable
  any more.   Thus the beam  propagates in preferentially  low density
  regions  with  rare encounters  of  gravitationally collapsed,  high
  density patches (halos) resulting in highly inhomogeneous geometry.}

By  contrast, the  standard  approach implicitly  uses a  perturbative
analysis  and  a  fluid  continuum  model  by  treating  the  beam  as
propagating  in the background  and perturbed  FL geometry.  From this
viewpoint, it  is surprising that  the standard analysis of  SNIa data
leads to  a consistent result,  in particular with  other cosmological
probes.  Is a  smooth cosmological  model  a good  description of  our
universe, and in particular for interpreting data such as SNIa? If so,
can we understand clearly why this is the case?

Standard  perturbation  theory  reveals  there are  problems.  If  the
angular diameter distance as a  function of redshift in a perturbed FL
model differs from the  background by $\kappa(\beta)$ where $\beta$ is
the angular scale of observation (see below for definitions), then the
variance of $\kappa$ scales as,
\begin{equation}
\langle\kappa^2(\beta)\rangle^{1/2}\sim10^{-2}\sigma_8\Omega_0^{0.75}
z_s^{0.8}\left(\frac{\beta}{1^{\rm o}}\right)^{-1-{n}/{2}},
\label{kapbet}
\end{equation}
as derived in \cite{fb97, Mellier:1998pk}, using linear perturbation
theory ($n$  is the  spectral index of  the power spectrum  of density
fluctuations and $z_s$ the redshift of the source).  This estimate was
confirmed in  \cite{js97} considering  the nonlinear evolution  of the
power   spectrum,   but   on   scales  below   10   arcmin,   $\langle
\kappa^2(\beta) \rangle$  increases more steeply  than the theoretical
expectation of the linear theory and is 2 to 3 times higher. Note also
that~(\ref{kapbet}) implies that $\langle\kappa^2(\beta)\rangle^{1/2}$
becomes  of  order  $2\times10^{-2}$  on  an angular  scale  of  order
1~arcmin, so that  the variance becomes much larger  than unity on the
typical  angular size  of the  ray  bundle for  SNIa.  So  large-scale
structure induces  a stochastic dispersion of  the luminosity distance
which  is difficult to  quantify with  standard techniques  for narrow
beams.   As  soon   as  one   goes  down   to  much   smaller  scales,
inhomogeneities  also   induce  a  systematic  shift   away  from  the
background, since one  cannot neglect the higher order  terms. This is
much more serious.  The magnification  of a source behaves as $\mu\sim
1+2\kappa+3\kappa^2+\vert\gamma\vert^2+\ldots$, where  $\gamma$ is the
shear  of  the  source  (defined  below),  and  so  the  mean  of  the
magnification       is       $\langle\mu\rangle\sim1+\langle3\kappa^2+
\vert\gamma\vert^2\rangle+\ldots\not=1$.  Thus,  if  the  variance  is
large  on small  angular scales,  we expect  the overall  shift  to be
significant~-- potentially of order unity or larger~-- on these scales
too. Modelling this shift accurately is critical for interpreting SNIa
observations correctly.

To estimate the  effect of the inhomogeneities on  smaller scales, one
needs   to    provide   a    better   description   of    the   matter
distribution. Attempts  to include  a uniform component,  high density
halos and low  density zones (filaments and voids)  have been proposed
\cite{KM:2009,KM:2010}, but none go down to the required scale.

{ Narrow light bundles  travel large distances ($\gtrsim 100\hMpc$)
  with a  very low  probability of encountering  dark matter  halos of
  substantial  mass, which we quantify in the following section.   The  cuspy   density  profiles  of   the  halos
  additionally  reduce  the  probability  of  a bundle  to  cross  the
  central, high density regions.}  Thus the bundles are subject mainly
to Weyl focussing  (i.e. induced by the gradient  of the gravitational
potential).  These  lightrays are expected to  be demagnified compared
to   lightrays  propagating  in   a  FL   spacetime  of   mean  matter
density. When one  averages such ray bundles over  the whole sky, this
dimming is compensated by a small number of ray bundles that encounter
very large density inhomogeneities
,  which then results  in high  focussing for  those directions  and a
magnification.   The general  averaging  argument was  put forward  by
Weinberg \cite{weinberg}.

{A similar argument can be employed to determine the average
  density  encountered  by  a  light  bundle  from  a  supernova.  The
  probability of  a light bundle encountering  massive halos decreases
  with   halo   mass.    Massive   galaxies,   groups   and   clusters
  ($\gtrsim10^{12}\hMsol$) are rarely  encountered, yet, they comprise
  $\sim50\%$ of total mass within the universe.  }
Thus, if we measure SNIa in  typical directions in the sky, we observe
them  in directions  where the  density of  matter encountered  by the
relevant ray bundles may be  expected to be less than the cosmological
average (for almost all directions  are of this nature). This argument
does  not  apply  to  the  much  larger  angular  scales  relevant  to
measurements  of the  BAO and  CMB  peaks.  These  beams do  encounter
sufficient matter to on average correspond to the overall cosmological
density, as  argued by  \cite{weinberg}.  In these  circumstances, one
expects not only an extra dispersion in the Hubble diagram, induced by
the  spatial  inhomogeneity of  the  intervening  medium,  but also  a
systematic shift, induced by  an observational selection effect, which
may well be significant.

The goal  of this  article is to  investigate these effects.  We first
discuss modelling  the matter distribution in the  real universe. Then
we  consider the  general relativistic  problem  of light  rays in  an
arbitrary  spacetime, with  a  focus on  how  a light  beam reacts  to
inhomogeneities compared to a smooth spacetime.  We then consider some
different approximations used to model narrow beams, such as perturbed
FL models,  the Dyer-Roeder approximation,  as well as  presenting two
new  approximations. Finally,  we  consider the  problem  of how  Weyl
focussing  by many  point sources  is converted  into  Ricci focussing
associated with a smooth matter distribution.

\section{The matter distribution}
\begin{figure}
\includegraphics[width=\hsize]{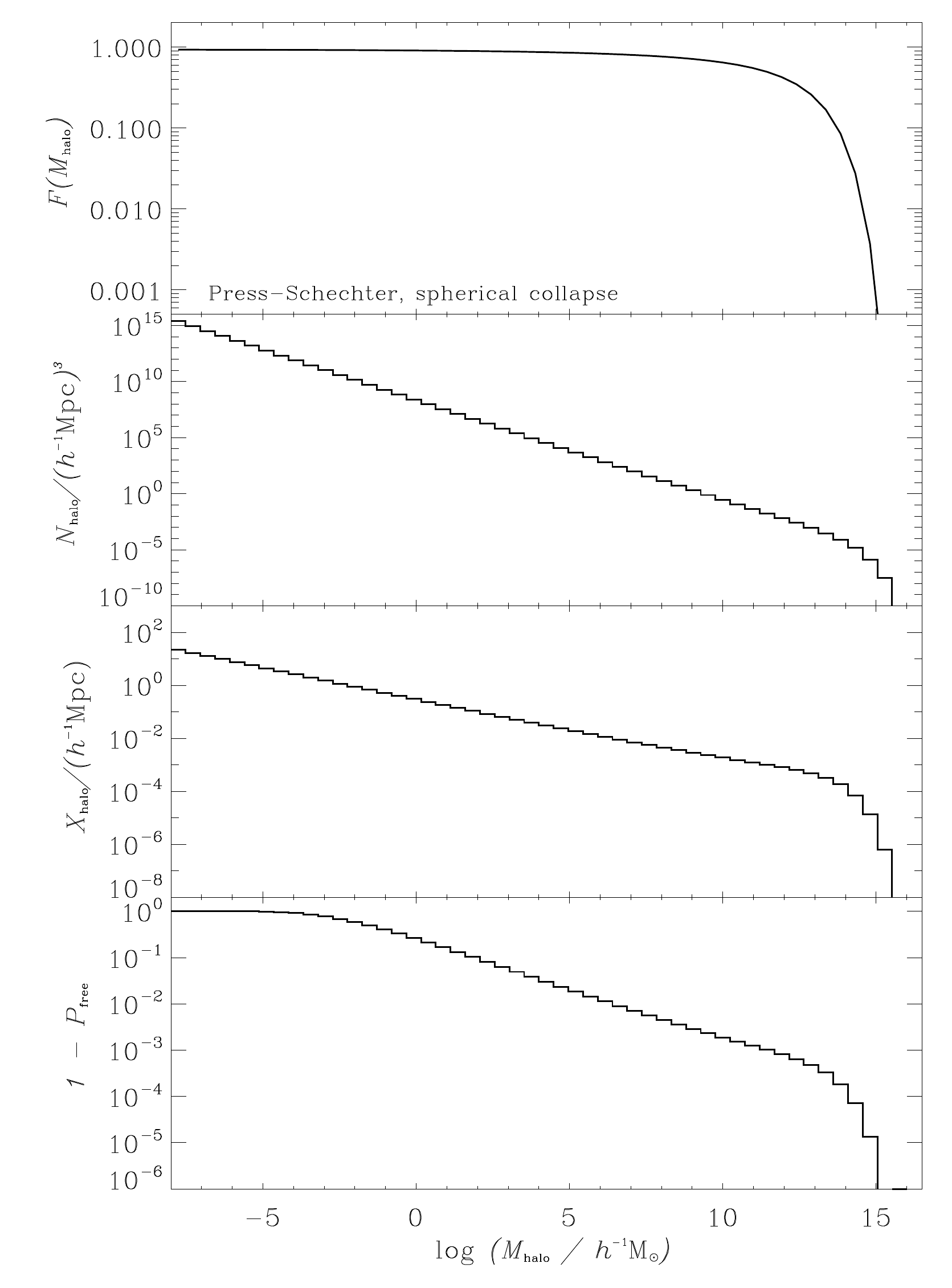}
    \caption{
      \label{fig:probX}  
      \normalsize
      {
        The top panel shows the total fraction of cosmic mass which is
        locked in  halos above a  given mass.  The Press-Schechter model predicts  50\% of  the total  mass is  locked in  halos above
        $3\times10^{11}\hMsol$. The  second panel gives  the number of
        halos per mass bin per  $\hMpcc$. The third panel displays the
        average  number  of  halos  per  mass bin  encountered  by  an
        infinitesimal  thin light  ray  per unit  length.  The  bottom
        panel indicates the probability  that a ray intersects with at
        least  one halo  within  a given  mass  bin on  a distance  of
        $1\hMpc$, which is equivalent  to $1- P_{\rm free}$, where $
        P_{\rm free}$ is the probability of not hitting a halo within that
        mass bin on a length of $1\hMpc$.}
    } 
\end{figure}
According to  the most accepted variants of  the $\Lambda$CDM paradigm
gravitationally  collapsed  structures span  a  mass  range from  
$10^{-8}\Msol$ (determined by the free-streaming scale of the CDM particle) to  $10^{15}\Msol$. In addition to the  matter bound in
collapsed structures a smooth component is expected which is not bound
to any structure. The question we are interested in here is, how these
structures affect the propagation  of light bundles travelling through
this inhomogeneous universe.

As  first attempt to  answer to  this question  we employ  an analytic
approach     based     on     a     Press-Schechter     (PS)     model
\citep{PressSchechter1974}. Press and  Schechter derived an analytical
expression for  the cumulative mass function,  $F(M_{\rm halo})$, which
gives  the   fraction  of   mass  locked  in   halos  above   a  given
mass. Integrating  their formula over  the whole mass  spectrum yields
one.  Thus  the PS model  predicts that {\em  all} matter is  bound in
gravitationally  collapsed halos,  with  masses ranging  from zero  to
infinity.  With the increasing  dynamical range of N-body simulations,
quantitative  differences   with  this  model   have  become  apparent
\cite{JenkinsEtAL2001,   ReedEtAl2003,  WarrenEtAl2006,  ReedEtAl2007,
  TinkerEtAl2008,     AnderhaldenDeiemnd2011,    FaltenbacherEtAl2010,
  MoreEtAl2011}. However, the difference between analytical and N-body
predictions depends  on the definition  of a halo in  the simulations.
Recently,  \cite{PradaEtAl2006,   CuestaEtAl2008}  showed  that  using
`dynamical masses' for N-body halos yields good agreement at least for
low redshifts.

The  top  panel  of  Fig.~\ref{fig:probX}  shows our computation of  $F(M_{\rm  halo})$
 based  on the 5-Year WMAP data
\cite{KomatsuEtAl2009}.  The  derivative of $F$  determines the number
density  of halos,  $N_{\rm  halo}$,  as a  function  of mass  (second
panel). According  to spherical collapse  theory the radius of  a halo
with mass, $M_{\rm halo}$, is given by:
\begin{equation}
r_{\rm halo} = \left( {3 M_{\rm halo}\over 4\pi\Delta_c \rho_{\rm crit}}\right)^{1/3}
\end{equation}
where  $\rho_{\rm crit}$  is the  critical  density  of  the universe  and
$\Delta_c = 95$ (for cosmological parameters chosen above).  

The
average  number of  halos, $X_{\rm  halo}$, with  mass  $M_{\rm halo}$
encountered by a single infinitesimal ray per unit length is the product of $N_{\rm
  halo}$ and the surface area of the halo.
\begin{equation}
X_{\rm halo} = N_{\rm halo} \pi r_{\rm halo}^2
\end{equation}
$X_{\rm halo}$ is  shown in the third panel from  the top. The average
number  of halos with  masses about  $10^{12}\hMsol$ encountered  by a
light ray is  $\sim0.001$. Thus, on average a single light ray
of light hits one $10^{12}\hMsol$ halo every $1000\hMpc$.  

The bottom panel  displays the probability that a  ray intersects with
at least one 
halo of a given mass while travelling a distance of $1\hMpc$, which is
equivalent  to   $1-P_{\rm  free}$,   where  $P_{\rm  free}$   is  the
probability of  passing $1\hMpc$ freely, i.e.,  without encountering a
single halo of that mass.   To compute $P_{\rm free}$ we subdivide the
volume into  cubes with $1/N_{\rm  1D}=N_{\rm halo}^{-1/3}$ on  a side
and  assume  that  halos  within  a given  mass  bin  are  distributed
\emph{quasi homogeneously}, i.e, only one single halo is placed within
each  cube.  The  position of  the  halo  within  the cube  is  chosen
randomly.  With  these assumptions $P_{\rm free}$  can be approximated
by
\begin{equation}
P_{\rm free} = \left(1 - \pi r_{\rm halo}^2 N_{\rm 1D}^2\right)^{N_{\rm 1D}}\,.
\end{equation}
We conclude that a light  ray travelling $1\hMpc$ through the present,
nonlinear,  cosmic   density  field  encounters   with  almost  100\%
certainty several halos with  masses below $10^{-3}\hMsol$.  Since the
PS model does not include a  smooth component the predictions for this
mass  range must  be corrected  accordingly.  On the  other side,  the
probability to hit  at least one $10^{12}\hMsol$ halo  is of the order
of 0.1\%. These  results suggest we must discuss the  effects of small and
large scale structure on the light propagation separately.

\subsection{Effects of small scale structure}

Lensing can discriminate between a diffuse and smooth
component (a gas of microscopic particles) and one of
macroscopic massive objects (gravitationally bound),
and has been used~\cite{Metcalf1999,Metcalf2006} to probe the
nature of dark matter on galactic scales. The two
components can be characterized by a mass scale, defined by the fact
that the projected density be smooth on a scale of order the
angular size of the source. This gives~\cite{Metcalf1999}
\begin{equation}
M_*
\sim 2\times 10^{-23} M_\odot h^2 \left(\frac{\lambda_{\rm s}}{1
{\rm AU}}\right)^3,
\end{equation}
where $\lambda_{\rm s}$ is the physical size of
the source.

Another important mass scale is set by the requirement
that the angular size of the source, $\beta=\lambda_{\rm
s}/D_A(z_{\rm s})$, is smaller than the Einstein angular radius
$\theta_{\rm E}$  so that it can be considered as a true point
source~\cite{Metcalf1999}:
\begin{equation}
M_{\rm point}\sim 5\times 10^{-7}
M_\odot \left(\frac{\lambda_{\rm s}}{1 {\rm AU}}\right)^2
 \left(\frac{10^3 {\rm Mpc}}{D_A(z_s)}\right).
\end{equation}
If $M<M_*$, the component can be considered as diffuse on the scale of
the ray bundle. If $M_*<M<M_{\rm point}$ there
will be very few high
magnification events with most of the lines of sight being demagnified, according to the standard lensing paradigm (see below). These
two components affect the probability distribution function for the magnification. In the
extreme case where the matter is composed only of macroscopic point-like
objects, then most high-redshift SNIa would appear less bright
than in a universe with the same density distributed smoothly, with some very
rare events of magnified SNIa~\cite{Metcalf1999,rauch,HW97}.
It has been argued~\cite{sh99} that the magnification
of high-redshift SNIa can be a powerful discriminator
of the nature of dark matter. In particular, based
on numerical simulations, a few hundred
SNIa at $z\sim1$ could allow a 20\% determination of the
fraction of matter in compact objects.

The dispersion of SNIa data due to lensing has been estimated in
various ways, but a complete analysis may require us to go down to
scales where our knowledge of the distribution of matter is very poor.  
In particular, it is important to know the amount of diffuse matter
compared to the amount of matter in small compact halos. Knowledge of
the spatial distribution of the halos is required to determine the
dispersion of the observations, keeping in mind that one also expects
a bias since most lines of sigh are demagnified.  

The minimum mass of gravitationally bound structures is determined by  the nature  of  the CDM particle
itself.  Its  mass induces a free streaming scale which in turn gives
rise to a mass threshold below  which no  gravitationally bound
structure  can form (in contrast to the original PS approach where
there is no such cut off mass).   Currently,  neutralinos  are  the
most promising   CDM candidates. The lightest neutralino, with $m\sim
100\,$GeV,  is favoured, as  it  is both  weakly interacting  and  stable   (e.g.
\cite{BertoneEtAl2005}).  Its  free-streaming scale is $\sim
0.7\,$pc, with a corresponding  minimum halo  mass $M_{\rm fs}\sim10^{-8}\rm M_\odot$. 

In  principle,  cosmological N-body  simulations  can  be employed  to
determine  the  total  amount of  mass  in  the  smooth and  the  halo
component.   Knowing the  fundamental  properties of  the CDM
particle, the initial  conditions can be derived and  propagated to
the current epoch. However, the dynamical range required for this approach
exceeds current computational resources.

One  strategy  to  circumvent  the computational limitations  is  to
enclose a  small region  of very high  resolution within a  larger but
lower  resolution simulation . 
Using  this technique \cite{DiemandEtAl2005} found
that at $z=26$ the mass function  is steep, $\dd n(M)/\dd M \propto M^{2}$
down to  $M_{\rm fs}$. At that  time about 5\%  of the mass in  the high
resolution  region  has  collapsed  into gravitationally  bound  halos
(see Fig.  3 in   \cite{DiemandEtAl2005}).
Due to technical limitations this simulation has not been run further
than $z\approx26$.

Another strategy  to determine  the total mass  locked in halos  is
via excursion set  theory  \cite{Zentner2005},  which propagates
density perturbations stochastically  to generate  the halo  mass functions.
For a  $\Lambda$CDM model with $100\,$GeV  neutralinos,  75--80\%  of matter  is
locked in halos  at $z\lesssim 1$ \cite{AnguloWhite2010}.  The  remaining 20-25\% is
smoothly  distributed  without  being  associated with  any  collapsed structure.
\subsection{Effects of large scale structure}
Current N-body simulations provide a very reliable picture of the
cosmic large scale structure. However small scale structure can only
be resolved down to the given mass resolution limit of these
simulations (currently $\sim10^8\Msol$). Analytical approaches, like
those based on PS models, are not affected by mass resolution issues
but are much more sketchy by nature. In the following we will investigate both approaches.  
\begin{figure}
\includegraphics[width=\hsize]{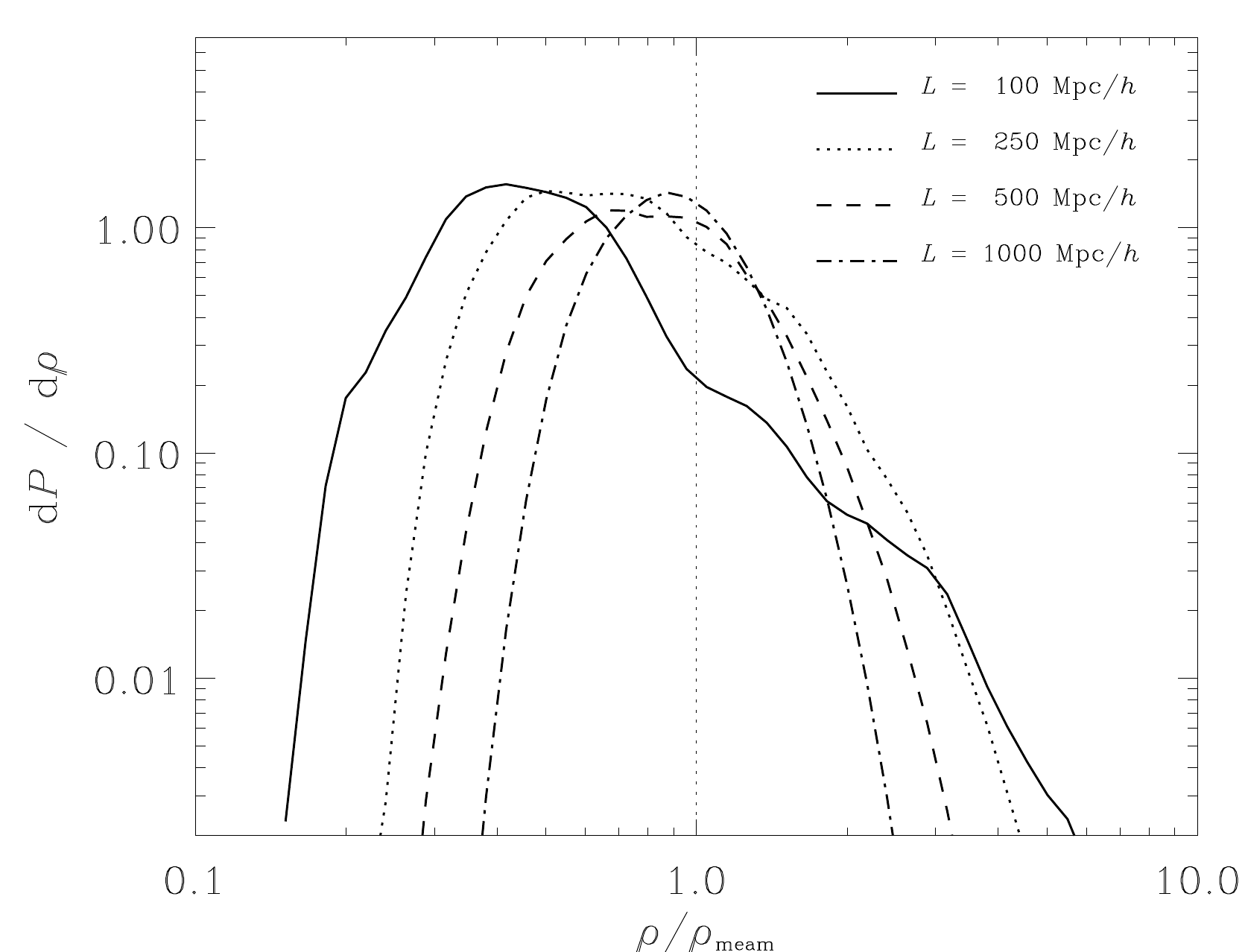}
    \caption{
      \label{fig:lidi}  
      \normalsize
      {Probability distribution of the averaged densities encountered
        by a single light ray of infinitesimal width for different path lengths based on a
        Press-Schechter model.}
} 
\end{figure}
\subsubsection{Analytical approach}
Based on the PS model discussed above one can determine the
probability distribution (PDF) of densities averaged along 
infinitesimally thin light rays. For that purpose we model the mass
distribution of halos by four bins with average masses from
$10^6\hMsol$ to $3\times10^{14}\hMsol$. The mass bins are chosen in such a way that the one dimensional
number density, $N_{\rm ÊÊ1D}$, increases by a factor of 10 for
each subsequent (decreasing) mass bin. The remaining mass contained
within halos below the smallest mass bin is assumed to be homogeneously
distributed. According to the PS model 15\% of the total gravitational 
matter is found in halos below $10^6\hMsol$. 

The probability of a ray encountering a given total number of halos  
is computed as product of the various binomial coefficients and
probabilities for hitting the partial number of halos per mass bin.
 
This approach allows us to compute the PDFs as a function of the path
length as shown in Fig.~\ref{fig:lidi}.  The distribution for a path length of 
$100\hMpc$ peaks at 0.4 times the mean density, $\rho_{\rm
  mean}$. For longer path lengths the peak shifts towards
$\rho_{\rm mean}$. But even for a path length of $1000\hMpc$
(corresponding to a redshift of $z\simeq0.25$) the distribution
peaks significantly below $\rho_{\rm mean}$. It is worth noting  
that by construction the integral of $\rho\ {\rm d}P$ from zero
to infinity equals unity. A peak below $\rho_{\rm mean}$ requires
a high density tail for counterbalance. 

The model presented here does not include halo clustering (inherent to
such kind  of approaches)  and assumes the  mass contained  in objects
below $10^6\hMsol$ to be distributed smoothly (to reduce computational
cost).  These   shortcomings  let  us  abstain   from  a  quantitative
interpretation at this point. But we can clearly see that the majority
of light  rays encounters averaged  densities below the mean  and that
the shape of the PDFs is a function of path length.
\begin{figure*}
\includegraphics[width=\hsize]{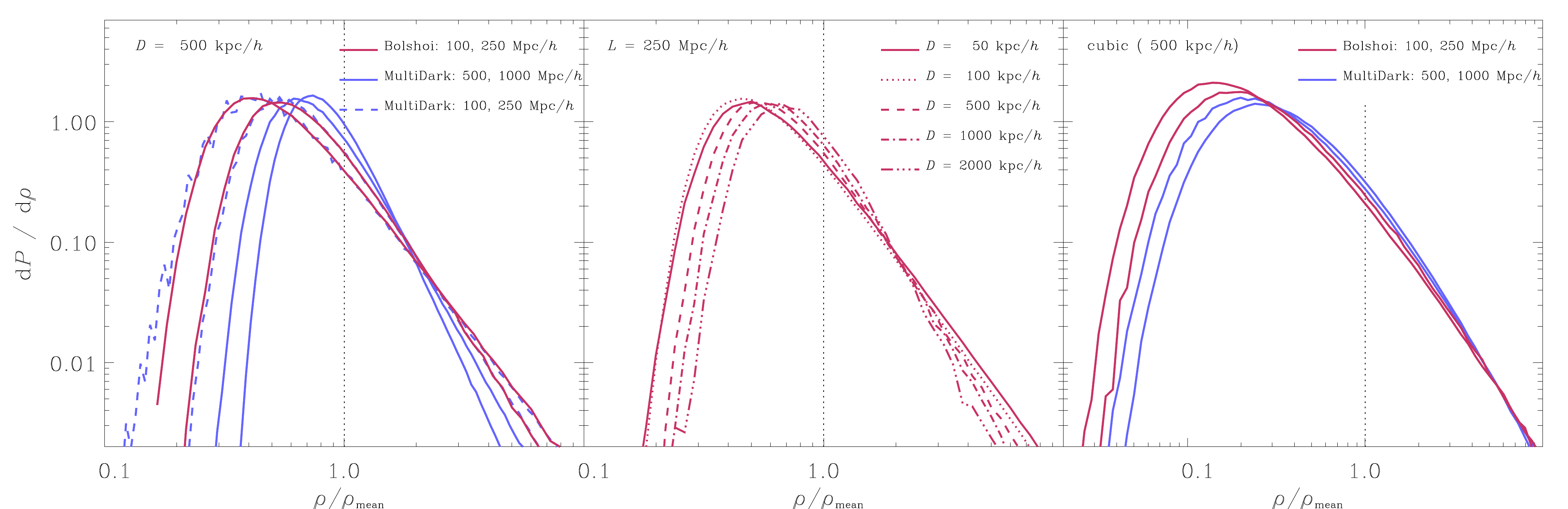}
    \caption{
      \label{fig:MultiDark}  
      \normalsize
      {Left panel: Probability distribution of averaged densities
        within long narrow ($0.5\hMpc$ diameter) beams for different
        path lengths using different simulations (indicated). 
        For short path lengths (e.g., $\sim100\hMpc$) most beams
        encounter low densities which is 
        counterbalanced by comparatively few beams which encounter much higher
        densities. With increasing path length the peak of the PDF
        approaches the cosmic mean, but even for a beam length of
        $1 \hGpc$ the peak occurs significantly below the mean density. 
        Independent of length the average 
        density encountered by a sufficiently large number of beams is equal to the
        cosmic mean density, i.e., the density weighted integrals
        for all curves shown above yields cosmic mean density. 
        Middle panel: PDF for beams of the same length ($250\hMpc$)
        but different diameter (indicated). As the
        beam becomes narrower the PDF broadens and the location of the
        peak tends to shift to slightly smaller densities.
        Right panel: Same as left panel except here the density is measured
        within cubes which have the same volume as the long and narrow
        beams. Note the striking difference between the PDFs: the peak position and widths change when comparing tubes to cubes, and the power law tail which is prominent for beams is no longer present. 
      }
    } 
\end{figure*}
\subsubsection{Numerical approach}
Numerical simulations provide detailed insight into the nonlinear
matter evolution inaccessible to analytical descriptions. However,
they are inevitably limited in their mass resolution. 
The resolution is proportional to the simulation volume
and inversely  proportional the number of phase space elements
(particles) used. For cosmological applications generally large volumes
are desirable. The number of particles is limited by the available
computational power. Particles of current state-of-the-art
cosmological simulations have masses of a few times $10^8\Msol$.
For comparison, the total mass, assuming homogeneous density distribution,
contained within the volume of a light beam of 1 AU diameter and
$1000\Mpc$  length (corresponding to $z\approx0.25$) is 
$\sim10^{-9}\Msol$.

In this section we compute the PDFs of the mean densities
within long and narrow `beams' based on a set of publicly available
N-body simulations, namely the Bolshoi \citep{Klypin:2011}, the
Millennium \citep{Springel:2005} and the MultiDark R1
\citep{Prada:2011} Simulation. (Descriptions of the data bases are given in~\cite{Lemson:2006,Riebe:2011}). These simulations 
compute the CDM distribution within cubes of $250^3$,
$500^3$ and $1000^3 \hMpcc$ with mass resolution of $1.3 \times
10^8$,  $8.6 \times 10^8$ and $8.7 \times 10^9\hMsol$, respectively. 
These mass resolutions only allow the determination of the densities
within beams wider than a few tens of kpc  
which is many orders of magnitude larger than the expected diameter of a
light beam from a supernova. Nevertheless the results derived here can
give basic insight into the mean densities within the volume of the
light beams from distant SNIa.

The right panel of Fig.~\ref{fig:MultiDark} shows the PDFs of the
mean densities within beams of $500\hkpc$ 
diameter as a function of their length. The dashed lines based on the
MultiDark simulation are shown as a consistency check. They are
expected to coincide with the Bolshoi results but are more affected by
Poisson noise due to the lower mass resolution. The shape of the
distributions is similar to those based on the PS model
(Fig.~\ref{fig:lidi}). Independent of length all distributions peak
below the mean density. The cumulative probability for a mean 
density below the cosmic mean for the 100, 250, 500 and
1000$\hMpc$ beams is 75\%, 71\%, 68\% and 65\%, respectively. The
large probability of an averaged density below cosmic mean is
counterbalanced by unlikely high density encounters. 
Investigations based on the Millennium simulation confirm these
results. The middle panel of Fig.~\ref{fig:MultiDark} shows the PDFs
for different diameters. There may be an indication that with decreasing
diameters the location of the peak converges towards values close to
$0.5 \times \rho_{\rm   mean}$, but we are unable to probe beam sizes below $\sim50\hkpc$. The shape of the distribution, and in particular that the location
of the peak is below unity, is preserved independent of diameter and
is expected to hold for much smaller diameters. The right hand panel of
Fig.~\ref{fig:MultiDark} displays 
the PDFs based on cubic volumes, i.e. the densities are measured in cubes
rather than beams (left panel) while the volume remains the same. The
overall shape of the PDFs is very similar to those shown on the left
but the location of the peak is shifted towards smaller densities. 
Obviously, the geometry of the `test volumes' has an impact on the
PDFs. The PDFs can not be accurately determined without incorporating
their spatial information of the large scale density distribution.  

The picture arising from the N-body simulations is consistent
with the PS model. The probability that a light beam from a supernova
encounters an average density less then cosmic mean is larger than 
50\%. The exact value depends on light path length. 
With shorter distances the cumulative probability of sampling
densities below cosmic mean increases. This effect may be sufficient to
induce biases in the luminosity distance relation.

\section{Light Propagation}

From a theoretical point of view, the effects of matter
inhomogeneities can be described by the geodesic deviation equation,
which describes the evolution of a bundle of geodesics
$x^\mu(v,s)$, where $v$ is the affine parameter and $s$ labels the geodesics. The past lightcones of the central observer are given by $w =\,$const, where $w$ is the phase. Then $k_\mu = \partial_\mu w$, so these curves are irrotational null geodesics:
\begin{equation}
k^\mu k_\mu = 0, \,\, k^\nu \nabla_\nu k_\mu = 0, \,\, \nabla_{[\mu}k_{\nu]}=0.
\label{nullgeod}
\end{equation}

The connecting vector $\eta^\mu = \dd x^\mu/\dd s$ relates
neighbouring geodesics with tangent vector
$ k^\mu = {\dd x^\mu}/{\dd v}$ to an arbitrary reference geodesic of the
bundle, $\bar x^\mu(v)=x^\mu(v,0)$, giving the
distance between neighbouring geodesics and hence the physical size and shape
of the bundle as one follows it down into the past. The connecting vector can always be
chosen such that $k^\mu\eta_\mu=0$ and it evolves according to the
geodesic deviation equation:
\begin{equation}\label{gde1}
 k^\alpha k^\beta \nabla_\alpha \nabla_\beta \eta^\mu = {R^\mu}_{\nu\alpha\beta} k^\nu k^\alpha\eta^\beta.
\end{equation}
This equation describes the change of shape of the bundle.

For fundamental observers with four-velocity
$u^\mu$ ($u^\mu u_\mu=-1$),  the redshift is defined by
\begin{equation}\label{blah}
 1+z(v) =\frac{(k_\mu u^\mu)_{v}}{(k_\mu u^\mu)_{0}},
\end{equation}
where the past-directed photon four-momentum is
\begin{equation}\label{dec-k}
 k^\mu =  (1+z)(-u^\mu + e^\mu), ~~e^\mu u_\mu =0 ,\,e^\mu e_\mu=1.
\end{equation}
Here $e^\mu $ is the spatial direction of observation,
and the spatial direction of propagation is $n^\mu= -e^\mu$.
The affine parameter increases
monotonically along each ray and coincides in an infinitesimal
neighborhood of the observation point with the Euclidean distance in
the rest frame of $u^\mu(0)$. Note that while it depends on the
4-velocity $u^\mu(0)$ of the observer, it does not depend on the
4-velocity $u^\mu(\bar x^\alpha(s))$ of the observed source.

The screen space at each point along a ray is in the observer's rest space and orthogonal to the ray direction. It is spanned by unit vectors $n^\mu_a$ ($a=1,2$), with
$g_{\mu\nu}n^\mu_a n_b^\mu =\delta_{ab}$ and $n_a^\mu u_\mu =
n^\mu_a k_\mu =0$, that are parallel transported along the
ray ($k^\mu\nabla_\mu
n_a^\nu =0$). We can choose the connecting vector to lie in the screen space,
so that\footnote{This is the Sachs basis, unique up to transformations
$n_a^\mu \rightarrow r_{ab}(\alpha) n_b^\mu + p_a k^\mu$, where $r_{ab}(\alpha)$ is a rotation through angle $\alpha$, and $p_a$ are constants.}
$ \eta^\mu = \eta_1 n_1^\mu+ \eta_2 n_2^\mu.$
By (\ref{gde1})
\begin{equation}\label{gde3}
 \frac{\dd^2}{\dd v^2}\eta_a = {\cal R}_{ab}\eta^b,
\end{equation}
where ${\cal R}_{ab}={R}_{\mu\nu\alpha\beta}k^\nu k^\alpha n_a^\mu n_b^\beta$ is the screen projection of the Riemann tensor.
We write
\begin{equation}
 {\cal R}_{ab} =
  \left(\begin{array}{cc}\Phi_{00} & 0 \\ 0 & \Phi_{00}\end{array} \right)
  +
  \left(\begin{array}{cc} - {\rm Re}\,\Psi_0 & {\rm Im}\,\Psi_0 \\
  {\rm Im}\,\Psi_0 & {\rm Re}\,\Psi_0\end{array} \right)
\end{equation}
with
\begin{equation}\label{ricweyl}
 \Phi_{00}=-\frac12 R_{\mu\nu}k^\mu k^\nu,\quad
 \Psi_0=-\frac{1}{2}C_{\mu\nu\alpha\beta}m^\mu
 k^\nu m^\beta k^\beta,
\end{equation}
where $m^\mu \equiv n_1^{\mu} - {\rm i}  n_2^{\mu}$.
The Einstein equations give $R_{\mu\nu}k^\mu k^\nu=8\pi G T_{\mu\nu}k^\mu k^\nu$, where $T_{\mu\nu}$ is the total energy-momentum tensor,
\begin{equation}
T_{\mu\nu}=(\rho+p)u_\mu u_\nu + pg_{\mu\nu}+\pi_{\mu\nu}+ q_\mu u_\nu+ q_\nu u_\mu.
\end{equation}
Here $\pi_{\mu\nu}$ is the anisotropic stress and $q_\mu$ is the momentum density. (For a perfect fluid $\pi_{\mu\nu}=0=q_\mu$; for more general fluids, we can always choose $q_\mu=0$, corresponding to the frame where comoving observers see no momentum flux). Then we find
\begin{equation}\label{Phi00-matter}
 \Phi_{00}(v)= -4\pi G [1+z(v)]^2\left(\rho+p+2q_\mu e^\mu+\pi_{\mu\nu}e^\mu e^\nu\right)\bigg|_{\bar x^\alpha(v)}.
\end{equation}
Note that a cosmological constant $\Lambda$ makes {\em no} contribution to $\Phi_{00}$.

The linearity of (\ref{gde3}) implies that
\begin{equation}
 \eta^a(v) = {{\cal D}^a}_b(v) \left.\frac{\dd\eta^b}{\dd v}\right\vert_{v=0},
\end{equation}
where ${{\cal D}^a}_b$ is the Jacobi map. By (\ref{gde3}), we have
the Jacobi matrix equation
\begin{equation}\label{gde2}
 {\frac{\dd^2}{\dd v^2}{{\cal D}^a}_b={{\cal R}^a}_c{{\cal D}^c}_b},\quad \eta^a(0) =0,\quad
 \frac{\dd{{\cal D}^a}_b}{\dd v}(0)={\delta^a}_b.
\end{equation}
This second-order linear equation can be rewritten as a first-order
nonlinear equation:
\begin{equation}
 \frac{\dd}{\dd v}{{\cal S}^a}_b + {{\cal S}^a}_c {{\cal S}^c}_b = {{\cal R}^a}_b,
\end{equation}
by defining the deformation matrix
\begin{equation}\label{def-S}
 \frac{\dd}{\dd v}{{\cal D}^a}_b = {{\cal D}^a}_c {{\cal S}^c}_b.
\end{equation}
The Jacobi map ${{\cal D}^a}_b$ or equivalently the
deformation matrix ${{\cal S}^a}_b$ are the central quantities to
describe the distortion of the geodesic bundle.
The deformation matrix is usually decomposed as\footnote{Recall that the null rotation $\nabla_{[\mu}k_{\nu]}$ vanishes since $k_\mu=\partial_\mu w$.}
\begin{equation}\label{deform}
 {{\cal S}^a}_b =
   \left(\begin{array}{cc} \hat\theta & 0 \\ 0 & \hat\theta \end{array} \right)
+
  \left(\begin{array}{cc} \hat\sigma_{1} & \hat\sigma_{2} \\ \hat\sigma_{2} & -\hat\sigma_{1}\end{array} \right),
 \end{equation}
which defines the optical scalars
$\hat\theta$ (null expansion) and $\hat\sigma\equiv \hat\sigma_1
+ {\rm i} \hat\sigma_2$
(null shear).
These satisfy the Sachs equations \cite{sachs}
\begin{eqnarray}
 \frac{\dd\hat\theta}{\dd v} + \hat\theta^2 + \vert\hat\sigma\vert^2 &=& \Phi_{00}, \label{e.1}\\
 \frac{\dd\hat\sigma}{\dd v} +2 \hat\theta \hat\sigma &=& \Psi_0 ,\label{s2} \\
 \hat\theta \equiv {1\over2}\nabla_\mu k^\mu,~~|\hat\sigma^2| & \equiv & {1\over2}\nabla_\mu k_\nu \nabla^\mu k^\nu-\hat\theta^2. \label{s3}
\end{eqnarray}

The evolution of a ray
bundle can then be discussed in terms of {\em Ricci} focussing ($\Phi_{00}$)
and {\em Weyl} focussing ($\Psi_0$). The first is generated by matter inside
the beam [see (\ref{Phi00-matter})] while the second derives from matter outside the beam,
which can generate a non-vanishing Weyl tensor inside the beam.
This distinction leads to
the problem raised by Zel'dovich \cite{zel} and Feynman \cite{feyn}, and posed in terms of the curvature tensor by Bertotti \cite{Bertotti}:
if the matter of the universe is clustered
in massive galaxies, the bundle propagates almost exclusively in vacuum,
or at least in underdense regions, and is thus
mostly subject only to the Weyl focussing; by contrast, the cosmological effect is modelled using
a homogeneous fluid which generates only Ricci focussing (the Weyl tensor vanishes in FL spacetime).
Dyer and Roeder \cite{DR72} (see also \cite{DR73,DR74,DR81})
effectively reproduced Zel'dovich's idea and
proposed an ansatz to model the propagation in regions with no intergalactic medium. Weinberg \cite{weinberg} disputed this model, arguing that multiple Weyl deflections by individual masses average
to mimic the Ricci effect of a fluid with equal average density.
Weinberg's argument, based on photon flux conservation,
is effectively the basis for the standard perturbative approach -- i.e. that when averaged over distances and angles, the divergence from vacuum or underdense regions is compensated by the convergence due to clumping, so that the average luminosity distance is the same as the luminosity distance in the FL background (see e.g. \cite{fangwu}).
Weinberg's argument has been disputed \cite{Ellis:1998ha,Rose:2001qi}. Later work (e.g. \cite{Kibble:2004tm,Kostov:2009uc}) has not produced a definitive answer to the question, in particular for the case of the very narrow beams involved in SNIa observations.

In order to properly describe a thin geodesic bundle, we need
to have a good description of the matter distribution on the scales of the
extension of the bundle, and determine how the effect of the inhomogeneities
average during the propagation of the bundle, with two main issues in mind:
(1)~determining the typical amplitude of the effect and (2)~understanding why the description
by a smooth universe seems to provide a good description and determine
its validity.

\subsection{Angular distance}

The Jacobi matrix can be diagonalized by rotations:
\begin{equation}
 {{\cal D}^a}_b =
r(-\alpha_1)
  \left(\begin{array}{cc} D_+ & 0 \\ 0 & D_- \end{array} \right)
r(\alpha_2) ,
\end{equation}
where the shape parameters $D_\pm$ are nonzero almost everywhere. Their
absolute values give the semi-axes of the (elliptic) cross-section of the bundle.
Once $D_\pm$ are fixed, the angles $\alpha_{1,2}$ are unique at
all points where the bundle is non-circular.

The {area distance} or { angular diameter distance} is then defined as\footnote{Note that the terminology `angular diameter distance' has two interpretations: as an area angular diameter distance, as used here, and as a linear angular diameter distance which are $D_+$ and $D_-$~\cite{Tomita:1999tg}.}
\begin{equation}
 D_A(v) = \sqrt{{\rm det}\,|{\cal D}(v)|}= \sqrt{|D_+(v)D_-(v)|}.
\end{equation}
For a bundle converging at the observer, $D_A$ relates the cross-sectional area $A$ at the
source to the opening solid angle at the observer. It depends on the 4-velocity
of the observer, but not of the source.
From (\ref{s3}), the null convergence
is
\begin{equation}
\hat\theta = \frac{1}{\sqrt{A}}\frac{\dd}{\dd v} \sqrt{A},
\end{equation}
and (\ref{e.1}) becomes
\begin{eqnarray}
\frac{\dd^2 D_A}{\dd v^2} &=& - \left(\vert\hat\sigma\vert^2   - \Phi_{00}\right)D_A. \label{e125}
\end{eqnarray}
For
$T_{\mu\nu}k^\mu k^\nu>0$ we have $\Phi_{00}\leq 0$ by
(\ref{Phi00-matter}), so that $\vert\hat\sigma\vert^2 - \Phi_{00} \geq 0$.
Thus
\begin{eqnarray}
 \frac{\dd^2 D_A}{\dd v^2}\leq0,
\end{eqnarray}
in any cosmological model, as long as the null energy condition
holds, irrespective of the value of the cosmological constant.

In order to compare to observations,
we need the relation
between $v$ and $z$. We have
$\dd z/\dd v =
\dd(u^\mu k_\mu)/\dd v = k^\nu\nabla_\nu(u^\mu k_\mu)
  = k^\mu k^\nu\nabla_\nu u_\mu$.
Now
\begin{equation}\label{e-du}
  \nabla_\mu u_\nu= \frac13 \Theta \left(g_{\mu\nu}+u_\mu u_\nu\right) + \sigma_{\mu\nu} +\omega_{\mu\nu}-  u_\mu A_\nu,
\end{equation}
where  $\Theta$ is the expansion, $\sigma_{\mu\nu}$ is the shear, $ \omega_{\mu\nu}$ is the vorticity
and $A^\mu$ is the acceleration.
In a universe containing CDM and baryons with four-velocity $u^\mu$, and $\Lambda$ (where radiation is dynamically negligible), we have $A_\mu=0$.

By Eqs.~(\ref{blah}) and~(\ref{dec-k}), we obtain \cite{Clarkson:2010uz}
\begin{equation}\label{vofz}
\frac{\dd z}{\dd v} =
(1+z)^2\left(\frac13 \Theta + \sigma_{\mu\nu}e^\mu e^\nu - A_\mu e^\mu \right).
 \end{equation}
For any quantity $X$ evaluated along the ray bundle
\begin{equation}\label{vofz2}
\frac{\dd X}{\dd v} =
(1+z)^2H_\parallel(z,e^\mu)\frac{\dd X}{\dd z},
\end{equation}
where $H_\parallel$ is the observed expansion rate along the line of sight \cite{Clarkson:2010uz},
\begin{equation}\label{vofz2b}
 H_\parallel(z,e^\mu) = \frac13 \Theta + \sigma_{\mu\nu}e^\mu e^\nu - A_\mu e^\mu.
\end{equation}
The observed expansion rate is made up of an isotropic expansion monopole, an acceleration dipole and a shear quadrupole.

The set of equations~(\ref{s2}), (\ref{e125})  and~(\ref{vofz})
is the basis for analyzing the effect of inhomogeneities.
There are four physical effects induced by
inhomogeneities that need to be taken into account:
\begin{description}
\item[area distance modifications] due to the difference between
Ricci focussing (when the rays move through a uniform medium) and Weyl
focussing (due to the tidal effects of nearby matter);
\item[redshift adjustment] due to the
differences between the true redshift of a source
and its redshift in a smoothed out model;
\item[affine parameter distortions] since inhomogeneities change the relation $v(z)$ (this is actually where $\Lambda$ affects observational relations);
\item[displacement of the light beam] since the ray path is  shifted sideways by inhomogeneities
and so experiences different Weyl and Ricci terms at the same $v$ because it is at a different spacetime point.
\end{description}

\subsection{Links with weak lensing formalism}

The weak lensing amplification matrix ${\cal A}$ relates the direction of observation,
\begin{equation}
\theta^a \equiv \left.\frac{\dd \eta^a}{\dd v}\right|_0,
\end{equation}
to the direction of the source,
\begin{equation}
 \theta^a_S = \frac{\eta^a(v)}{\bar D_A(v)}={{\cal A}^a}_b\theta^b,
\end{equation}
so that
\begin{equation}\label{def-A}
 {{\cal A}^a}_b(v) = \frac{{{\cal D}^a}_b(v)}{\bar D_A(v)}.
\end{equation}
Here $\bar D_A$ is the angular distance
in a FL background, and ${{{\bar{\cal A}}^a}}_{~b} =\delta^a_b $.
We decompose ${\cal A}$ into a shear $(\gamma_1,\gamma_2)$, a convergence $\kappa$ {and a rotation $\omega$}, so that
\begin{equation}\label{ekappa}
 {{\cal A}^a}_b =
  \left(\begin{array}{cc}1-\kappa-\gamma_1 & \gamma_2-\omega \\ \gamma_2+\omega & 1-\kappa+\gamma_1\end{array} \right).
\end{equation}
The magnification is given by
\begin{equation}
\mu \equiv {S \over S_0}= {1\over (1-\kappa)^2 -\vert\gamma\vert^2+\omega^2},
\end{equation}
where $S, S_0$ are the surface areas of the image and source
($S=S_0/{{\rm det}{\cal A}}$). {Note that while ${\cal R}_{ab}$ is symmetric by construction (see
Eq.~(\ref{gde3})) and ${\cal S}_{ab}$ is also symmetric for a bundle converging at the observer, it is not necessarily the case for ${\cal
D}_{ab}$, which actually cannot be generically symmetric; see e.g. Eq.~(\ref{def-S}).} 

The amplification matrix  (\ref{def-A}) and the
deformation matrix (\ref{def-S}) are both related to the Jacobi matrix, and hence they are related by
\begin{equation}
 \bar D_A\frac{\dd}{\dd v}{\mathcal{A}^{a}}_{b}
       +{\mathcal{A}^{a}}_{b}\frac{\dd}{\dd v}\bar D_A=\bar D_A{\mathcal{A}^{a}}_{c}{\mathcal{S}^{c}}_{b}.
\end{equation}
This implies (away from caustics, where det$\,{\cal A}=0$),
\begin{equation}\label{LinkEq4}
{\left(\mathcal{A}^{-1}\right)^{a}}_{c}\left({\mathcal{A}^{c}}_{b}\right)'+{\delta^a_b}\frac{\bar D_A'}{\bar D_A}={\mathcal{S}^{a}}_{b},
\end{equation}
with a prime denoting $\partial/\partial v$ and where
\begin{equation}
{\left(\mathcal{A}^{-1}\right)^{a}}_{b} =
\mu\left( {\begin{array}{cc}
 1-\kappa +\gamma_1 &- \gamma_2+\omega \\
 - \gamma_2-\omega & 1-\kappa-\gamma_1 \\
 \end{array} } \right).
\end{equation}
{The Sachs optical scalars are then given by
\begin{eqnarray}
  \hat{\theta} &=&  \left(\ln\frac{D_A}{\sqrt{\mu}}\right)' 
  \label{ResEq3}\\
 \hat{\sigma}_1&=& -\mu\left[(1-\kappa)\gamma_1'+\gamma_1\kappa'+\gamma_2\omega'-\omega\gamma_2' \right] ,\label{ResEq1}\\
 \hat{\sigma}_2&=&  \mu\left[(1-\kappa)\gamma_2'+\gamma_2\kappa'-\gamma_1\omega'+\omega\gamma_1' \right], \label{ResEq2}
\end{eqnarray}
with the constraint that $\omega\gamma_2'-\gamma_2\omega'=(1-\kappa)\gamma_1'-\gamma_1\kappa'$ that arises from $\nabla_{[\mu}k_{\nu]}=0$.} 
These relations are useful since the optical scalars are more general
(they are defined for any spacetime geometry), while the weak lensing scalars are widely used in cosmology (but they assume a FL
background, see Ref.~\cite{ppu} for a general description). {While generically $\omega\not=0$, one can see that these equations
imply that for a FL spacetime only $\hat\theta=D_A'/D_A$ is
non-vanishing at the background level while the shear appears only at linear order in perturbation
and one has $(\hat\sigma_1,\hat\sigma_2)=(-\gamma_1,\gamma_2)$ and the
rotation appears only et second order in perturbations.}

\subsection{From affine parameter to redshift dependence}

The evolution equations (\ref{s2}), (\ref{e125}) for the null shear and angular distance are in terms of the unobservable affine parameter $v$.
We need to convert to the observed redshift, using (\ref{vofz}).
Using (\ref{vofz2}), we obtain
\begin{eqnarray}
\frac{\dd^2z}{\dd v^2}
&=&
k^{\mu}k^{\nu}k^{\alpha}\nabla_{\mu}\nabla_{\nu}u_{\alpha} \nonumber\\
&=&
 -\frac{2}{3}(1+z)^3H_\parallel\Theta
 -\frac{1}{3}
 (1+z)^3k^\alpha\nabla_\alpha\Theta
 \nonumber\\
 &+&(1+z) k^\mu k^\alpha\nabla_\alpha A_\nu
 +(1+z)^2H_\parallel k^\nu A_\nu
 \nonumber\\
 &-&
 k^\mu k^\nu k^\alpha\nabla_\alpha\sigma_{\mu\nu}.
\end{eqnarray}
The last term can be evaluated by expanding $k^\alpha$ with~(\ref{dec-k}) and using
$u^\nu u^\alpha\nabla\sigma_{\mu\nu}=-\sigma_{\mu\nu}A^\nu$
and $u^\nu e^\alpha\nabla\sigma_{\mu\nu}=-\sigma_{\mu\nu}e^\alpha\nabla_\alpha u^\nu$.
It follows that for any quantity $X$,
\begin{equation}
\frac{\dd^2X}{\dd v^2} = (1+z)^4H_\parallel^2\frac{\dd^2X}{\dd z^2}
+(1+z)^3Q \frac{\dd X}{\dd z},
\end{equation}
where
\begin{eqnarray}
Q&=&\frac23 \Theta H_\parallel - \frac13 \dot\Theta  +A_\mu A^\mu \nonumber\\&+&  e^\mu \Big(- \frac13\nabla_\mu\Theta +H_\parallel A_\mu  -\dot A_\mu - u^\nu\nabla_\mu A_\nu
+2\sigma_{\mu\nu}A^\nu \Big)
\nonumber \\ &+& e^\mu e^\nu\Big(\frac23\Theta\sigma_{\mu\nu} -\dot\sigma_{\mu\nu} -2{\sigma_\mu}^\alpha\sigma_{\alpha\nu} -2 {\omega_\mu}^\alpha\omega_{\alpha\nu} \nonumber\\
&+&\nabla_\mu A_\nu \Big)
-e^\alpha e^\nu e^\mu \nabla_\alpha \sigma_{\mu\nu}\,.\label{def-Q}
\end{eqnarray}

This form of $Q$ is completely general, for any spacetime geometry and energy-momentum tensor, and independent of the field equations. It is convenient to
write $H_\parallel^2$ and $Q$ in
terms of covariant multipoles, using
a covariant generalization of a spherical harmonic expansion  \cite{Clarkson:2010uz}.
We expand in terms of
the trace-free products $e^{\<\mu }e^{\nu\>}$, $e^{\<\mu}e^\nu e^{\alpha\>}$ and $e^{\<\mu}e^\nu e^{\alpha}e^{\beta\>}$, and use the spatial covariant derivative $\D_{\mu}$. Then we obtain \cite{cdmu}:
\begin{eqnarray}
H_{||}^2 &=& \frac{1}{9}\Theta^2 + \frac{1}{3}A_{\mu}A^{\mu}+ \frac{2}{15}\sigma_{\mu\nu}\sigma^{\mu\nu} \nonumber\\
&-& e^{\mu}\Big[\frac{2}{3}\Theta A_{\mu}+ \frac{4}{5} A^{\nu}\sigma_{\mu\nu}\Big] \nonumber\\
&+& e^{\<\mu}e^{\nu\>}\Big[ A_{\mu}A_{\nu}+ \frac{2}{3}\Theta \sigma_{\mu\nu}+ \frac{4}{7}\sigma^{\alpha}{}_{\mu}\sigma_{\nu\alpha}\Big] \nonumber\\
&-& 2e^{\<\mu} e^{\nu} e^{\alpha\>} A_{\mu}\sigma_{\nu \alpha}
+e^{\<\mu}e^{\nu}e^{\alpha}e^{\beta\>}\sigma_{\mu\nu}\sigma_{\alpha\beta},
\label{hpz}
\end{eqnarray}
and
\begin{eqnarray}
 Q&=&\frac{4\pi G}{3}(\rho+3p) -\frac{1}{3}\Lambda + \frac{1}{3}\Theta^2 + \sigma_{\mu \nu}\sigma^{ \mu \nu} \nonumber \\
 &-&
\frac{1}{3}\omega_{\mu\nu}\omega^{\mu\nu} + A_{\mu} A^{\mu}-\frac{2}{3}\D^\mu A_\mu
\nonumber \\
&+&
e^{\mu}\Bigg[\frac{1}{3}\D_{\mu}\Theta+\frac{2}{5}\D^\nu{\sigma}_{\mu\nu}
\nonumber\\
&+&
\dot A_{\mu} - \frac{4}{3} \Theta A_{\mu}- \frac{17}{5}A^{\nu}\sigma_{\mu \nu}
-A^{\nu}\omega_{\mu \nu}\Bigg]
\nonumber \\
&+&
e^{\<\mu }e^{\nu\>}\Big[E_{\mu \nu}-4\pi G\pi_{\mu \nu} + 2\Theta\sigma_{\mu \nu} + 3 {\sigma^{\alpha}}_{\mu}\sigma_{\nu \alpha} \nonumber\\
&+&
\omega_{\mu}{}^\alpha\omega_{\alpha\nu} +2\omega_{\alpha \mu} {\sigma_{\nu}}^\alpha - 2\D_{\mu}A_{\nu}\Big] \nonumber\\
&+&
e^{\<\mu}e^\nu e^{\alpha\>}\Big[\D_{\mu}\sigma_{\nu \alpha}- A_{\mu}\sigma_{\nu \alpha}\Big] \label{curlyQeqn1}
\end{eqnarray}
where we also used the covariant evolution and constraint equations of GR (see \cite{Tsagas:2007yx}).
Here $E_{\mu\nu}=C_{\mu\alpha\nu\beta}u^\alpha u^\beta$ is the electric part of the Weyl tensor (generalizing the Newtonian tidal tensor).
These expressions show clearly the covariant monopole and higher multipoles; for example, the octupole of $Q$ is $\D_{\<\mu}\sigma_{\nu \alpha\>}- A_{\<\mu}\sigma_{\nu \alpha\>}$. Note that the monopole of $H_\|^2$  has contributions from the shear even though the monopole of $H_\|$ does not.

Finally we can rewrite the evolution equation (\ref{e125}) for the angular distance in terms of redshift:
\begin{eqnarray}\nonumber
\hspace*{-8mm}&&\left(1+z\right)^2H_\parallel^2\frac{\dd^2D_A}{\dd z^2}+\left(1+z\right)Q\frac{\dd D_A}{\dd z}=\\
\hspace*{-8mm} &&\!-\!\left[\!4\pi G\Big(\rho+p+2q_\mu e^\mu+ \pi_{\mu\nu}e^\mu e^\nu\Big)+\frac{|\hat{\sigma}|^2}{\left(1+z\right)^2}\!\right]\! D_A .\label{zda}
\end{eqnarray}
This is a completely general and nonlinear equation, valid in any spacetime, with any matter content, where $H_\parallel^2$ and $Q$
are given by (\ref{hpz}) and (\ref{curlyQeqn1}).
The null shear terms are given by the remaining Sachs equation (\ref{s2}); in terms of redshift, this is
\begin{eqnarray}
&&H_\parallel\left(\frac{\dd\hat{\sigma}_a}{\dd z}+\frac{1}{D_A}\frac{\dd D_A}{dz}\hat\sigma_a\right)  \nonumber \\ &&
~~~~~~~ =-\left(E_{\mu\nu}-\varepsilon_{\mu\alpha\beta}e^{\alpha}
H_\nu{}^\beta\right)N_a^{\mu\nu}, \label{zsig}\\
&&N_a^{\mu\nu} \equiv \left(n^{\mu}_1n_{1}^{\nu}-n_{2}^{\mu}
n^{\nu}_{2},~n^{\mu}_1n_{2}^{\nu}+n_{2}^{\mu}
n^{\nu}_{1}\right),
\end{eqnarray}
where $H_{\mu\nu}={1\over2}\varepsilon_{\mu\alpha\beta}C^{\alpha\beta}{}_{\nu\kappa}
u^\kappa $ is the magnetic part of the Weyl tensor -- which has no Newtonian analogue -- and $\varepsilon_{\mu\nu\alpha}$ is the spatial alternating tensor.

Equations (\ref{zda}) and (\ref{zsig}) form a closed system that determines $D_A$ and $\hat\sigma$ in terms of $z$ and $e^\mu$.
In particular, we see what is required to determine $D_A$ in a lumpy universe:
the total energy-momentum tensor (i.e. $\rho,p,q_\mu, \pi_{\mu\nu}$), the kinematics of the fundamental four-velocity (i.e. $\Theta,\sigma_{\mu\nu}, \omega_{\mu\nu}, A_\mu$), the magnetic and electric part of the Weyl tensor, $H_{\mu\nu}$ and $E_{\mu\nu}$.

In a universe with dust matter (CDM and baryons, sharing the same four-velocity), with dark energy in the form of $\Lambda$ and where we can neglect radiation (i.e. at late times), we have
\begin{equation}
p=A_\mu=q_\mu= \pi_{\mu\nu}=0.
\end{equation}
From now on we will make this assumption, together with $\omega_{\mu\nu}=0$. Then $H_{||}^2$ and $Q$ simplify to
\begin{eqnarray}
H_{||}^2 &=& \frac{1}{9}\Theta^2 + \frac{2}{15}\sigma_{\mu\nu}\sigma^{\mu\nu}
+e^{\<\mu}e^{\nu\>}\Big[\frac{2}{3}\Theta \sigma_{\mu\nu}+ \frac{4}{7}\sigma^{\alpha}{}_{\mu}\sigma_{\nu\alpha}\Big] \nonumber\\ &+& e^{\<\mu}e^{\nu}e^{\alpha}e^{\beta\>}\sigma_{\mu\nu}\sigma_{\alpha\beta},
\label{hpz2}\\
 Q&=&\frac{4\pi G}{3}\rho -\frac{1}{3}\Lambda+ \frac{1}{3}\Theta^2   + \sigma_{\mu \nu}\sigma^{ \mu \nu}
\nonumber \\
&+&
e^{\mu}\Bigg[\frac{1}{3}\D_{\mu}\Theta+\frac{2}{5}\D^\nu{\sigma}_{\mu\nu}
\Bigg]
\nonumber \\
&+&
e^{\<\mu }e^{\nu\>}\Big[E_{\mu \nu} + 2\Theta\sigma_{\mu \nu} + 3 {\sigma^{\alpha}}_{\mu}\sigma_{\nu \alpha}
\Big]
\nonumber\\  &+&
e^{\<\mu}e^\nu e^{\alpha\>}\D_{\mu}\sigma_{\nu \alpha}, \label{Q3}
\end{eqnarray}
and the angular distance equation (\ref{zda}) becomes
\begin{eqnarray}
&&\left(1+z\right)^2H_\parallel^2\frac{\dd^2D_A}{\dd z^2}+\left(1+z\right)Q\frac{\dd D_A}{\dd z}\nonumber \\
 &&~~~~~~~~~ =-\left[4\pi G\rho+\frac{|\hat{\sigma}|^2}
{\left(1+z\right)^2}\right] D_A .\label{zda2}
\end{eqnarray}
The form of (\ref{zsig}) is unchanged.

\section{Models based on different approximations}

We briefly review the standard FL approach and the DR approximation, and then we propose and investigate modifications of the DR model. (For other related reviews, see also \cite{sasaki,Tomita:1999tg,Rasanen:2008be,Rasanen:2009uw}.)
The set of equations~(\ref{e.1}), (\ref{s2}) and~(\ref{vofz}) -- equivalently (\ref{zda}) and (\ref{zsig}) -- is completely general and
does not depend on the choice of a particular spacetime geometry. We show here
how they lead to different expressions for the angular distance as a function of redshift,
depending on the assumptions on the distribution of the matter.

\subsection{Smooth FL model}\label{par-fl}

If we assume the matter is smoothly distributed, then the universe
can be described by a FL geometry,
\begin{eqnarray}
\dd s^2 &=& a^2(\eta)\left[-\dd\eta^2 + \dd\chi^2 + f_K^2(\chi)\dd\Omega^2\right],\\
H(z)&=& H_0 \sqrt{\Omega_{\rm m0}(1+z)^3 + \Omega_{\Lambda0} + \Omega_{K0}(1+z)^2},
\label{e-fleq}
\end{eqnarray}
where $f_K(\chi)= \sin(\sqrt{K}\chi)/\sqrt{K}$
is the comoving angular distance.
The Weyl tensor vanishes, so that $\Psi_0=0$, and $\hat\sigma=0$ by (\ref{s3}), consistent with (\ref{s2}). Also,  ${\cal R}_a^b = \Phi_{00}\delta_a^b$, and $H_\parallel=H$ by (\ref{vofz2}).
Then using (\ref{Phi00-matter}), it follows that (\ref{zda}) reduces to
\begin{eqnarray}\label{dr-bgd}
\frac{\dd^2\bar D_A}{\dd z^2} &+& \left(\frac{\dd \ln H}{\dd z}
 +\frac{2}{1+z}\right) \frac{\dd\bar D_A}{\dd z}
 \nonumber\\
 &=& -\frac{3}{2}\Omega_{\rm m0}
\frac{H_0^2}{H^2}(1+z)
\bar D_A.
\end{eqnarray}
It is important to realize that $H(z)$
in this equation appears from the change of variable from $v$ to $z$.

This equation also follows directly from the Jacobi matrix
equation~(\ref{gde2}), which is easily solved after
a conformal transformation,
\begin{equation}\label{gdefl}
 \frac{\dd^2}{\dd\tilde v^2}{\cal D}^a_b = - K{\cal D}^a_b,
\end{equation}
where $\tilde v$ is the affine parameter in the conformal spacetime
of the static metric
$\dd \tilde s^2 = -\dd\eta^2 + f_K^2(\chi)\dd\Omega^2$.
The solution of (\ref{gdefl}) is ${\cal D}^a_b = f_K(\tilde v)\delta_b^a$.
One can choose $\tilde v$ either as $\eta$ or $\chi$ and the angular distance is then
given by the standard formula~\cite{pubook,schneiderbook,Perlick2004}
\begin{equation}\label{e-da0}
\bar D_A(z) = \frac{a_0}{(1+z)}f_K[\chi(z)].
\end{equation}
Along the past lightcone $\dd\chi=-{\dd\eta}$ and ${\dd z}/{\dd \eta}= -a_0H(z)$,
so that
\begin{equation}\label{chiz}
a_0H_0\chi(z) = \int_0^z\frac{\dd z'}{H(z')/H_0}.
\end{equation}
Since $\dd\chi=\dd z/H$,  (\ref{gdefl})
recovers  (\ref{dr-bgd}), after using
$\dot H = -(1+z) H\dd H/\dd z$
and the Einstein equation for $\dot H$.
We can either solve~(\ref{gdefl})  or~(\ref{dr-bgd})
but the first is more direct. Also, the
derivation of  (\ref{dr-bgd}) required the relation $v(z)$,
or equivalently $\tilde v(z)$.

For a FL universe ${\cal A}^a_b = \delta_b^a$
(i.e. $\kappa=\gamma_1=\gamma_2=0$) and then
${\cal S}^a_b = ({\dot f_K}/{f_K})\delta_b^a$, so that
only $\hat\theta=f_K^{-1}{\dd f_K}/{\dd\tilde v}$
is non-vanishing, which is a consequence of the spatial homogeneity
and isotropy. After integration of  (\ref{s2}),
$\sqrt{A}=f_K$ in the conformal spacetime, so that again we recover the same expression for the angular distance.

\subsection{Perturbed FL model}

The simplest way to account for
inhomogeneous matter is via perturbation theory.
At first order
for scalar perturbations,
\begin{equation}
 \dd s^2 = a^2(\eta)\left[-(1+2\Phi)\dd\eta^2 + (1-2\Psi)\gamma_{ij}\dd x^i \dd x^j \right],
\end{equation}
in Newtonian gauge, where $\Phi$ and $\Psi$ are the Bardeen potentials.
The angular distance is
\begin{equation}
 D_A= \bar D_A(1+\delta_A),
\end{equation}
and the distance duality relation implies that the luminosity distance is $D_L= (1+z)^2\bar D_A(1+\delta_A)$. Thus
\begin{equation}
 \delta_L(z,\bm{n})= \delta_A(z,\bm{n}) + 2\frac{\delta z(z,\bm{ n})}{(1+z)^2},
\end{equation}
where $\bm{n}$ is the direction
of observation. The second term encodes fluctuations of photon energy due to the local gravitational potentials as well as Doppler
effects, and
is similar to the Sachs-Wolfe effect on the CMB. It was investigated in \cite{Bonvin:2005ps}
and also estimated in \cite{ubm} in another context,
but neglected in \cite{Cooray:2005yp,Cooray:2005yr,Sarkar:2007sp,Cooray:2008qn,Dodelson:2005zt,
Vallinotto:2010qm},
which assumed $\delta_L= \delta_A$.

As long as the bundle remains in the weak lensing regime,
\begin{equation}
 \mu\simeq 1+2\kappa,
\end{equation}
so that $\delta_L(z,\bm{n})=-\kappa(z,\bm{n})$.
This assumes that the inhomogeneities can be described
by density fluctuations of a {\em homogeneous} field. Then $\delta_L$ is a stochastic field of zero mean, so that
$\langle\mu\rangle=1$ (in terms of ensemble average) and thus $\langle D_A(z,\bm{n})\rangle=\bar D_A(z)$.
In such a description, the FL angular distance is the
mean distance that a collection of observers will determine. There may
indeed be a bias from this prediction arising from our actual
position in the universe. This is a cosmic variance problem.

\subsubsection{Derivation from the Jacobi map equation}

The Jacobi equation~(\ref{gdefl}) reduces to
\begin{equation}\label{e.66}
 \frac{\dd^2}{\dd\tilde v^2} {\cal D}_{ab}^{(1)}+K {\cal D}_{ab}^{(1)}=
  f_K(\tilde v){\cal R}_{ab}^{(1)}(\tilde v),
\end{equation}
where ${\cal D}_{ab}={\cal D}_{ab}^{(0)}+{\cal D}_{ab}^{(1)}$,
and ${\cal D}_{ab}^{(0)}=f_K(\tilde v)\delta_{ab}$ is the background Jacobi
matrix derived above. This equation
has the integral solution
\begin{equation}
 {\cal D}_{ab}^{(1)}(\tilde v)= \int_0^{\tilde v}
 f_K(\tilde v')f_K(\tilde v-\tilde v'){\cal R}_{ab}^{(1)}(\tilde v')\dd\tilde v',
\end{equation}
so that the amplification matrix is given by
\begin{equation}
 {\cal A}_{ab}^{(1)}(v) = \int_0^v
 \frac{f_K(v')f_K(v-v')}{f_K(v)}
 {\mathcal{R}}_{ab}^{(1)}(v')\dd v',
\end{equation}
where $2 {\mathcal{R}}_{ab}^{(1)} =
\delta g_{\mu\nu,\alpha\beta}k^\mu k^\nu {n}_a^\alpha {n}_b^\beta$. Since we are interested in modes smaller than the Hubble scale, we
assume that the spatial curvature does not influence the perturbations and neglect it in the
computation of  ${\mathcal{R}}_{ab}^{(1)}$ while we keep it in the geometrical factors.
We find
${\mathcal{R}}_{ab}^{(1)} = -\partial_{a}\partial_{b}(\Phi + \Psi) = -2\partial_{a}\partial_{b}\Phi$, where $\Psi=\Phi$ since we can neglect anisotropic stress at late times and on sub-Hubble scales.
Then the amplification matrix is $\mathcal{A}_{ab}=\delta_{ab} -\partial_{a}\partial_{b}\psi(\bm{n},\chi)$ with
\begin{eqnarray}
\psi \equiv 2\int_0^\chi \frac{f_K(\chi')f_K(\chi-\chi')}{f_K(\chi)}
 \Phi[f_K(\chi')\bm{n},\chi']\dd\chi'.
\end{eqnarray}
We conclude that $\delta_A = -\kappa$ with
\begin{eqnarray}
&& \delta_A = -\kappa(\bm{n},\chi)=-\frac{3}{2}
H_0^2\Omega_{\rm m0}\times\nonumber\\
&&~~  \times \int\frac{f_K(\chi')f_K(\chi-\chi')}{f_K(\chi)}
   \frac{\delta[f_K(\chi')\bm{n},\chi']}{a(\chi')}\dd\chi',\label{e241}
\end{eqnarray}
where the Poisson equation has been used to replace $\nabla^2\Phi$ with $\delta$.
This gives the fluctuation of the angular distance in the direction $\bm{n}$,
taking into account propagation in a perturbed spacetime.

\subsubsection{Derivation from the Sachs equations}

The same result can, in principle, be derived from the Sachs equations~(\ref{e.1}), (\ref{s2}) and~(\ref{vofz}).
The background and first order equations are:
\begin{eqnarray}
 && D^{(0)\prime\prime}_A = \Phi_{00}^{(0)} D_A^{(0)},\label{fx1}\\
 &&z^{(0)\prime} = \frac13 \left[1+ z^{(0)}\right]^2\Theta^{(0)},\label{fx2}\\
 &&D^{(1)\prime\prime}_A  = \Phi_{00}^{(0)} D_A^{(1)} +   D_A^{(0)} \Phi_{00}^{(1)},\label{fx3}\\
 && z^{(1)\prime} = \frac23 \left[1+ z^{(0)}\right]\Theta^{(0)}z^{(1)}\label{fx4}\\
 &&  ~~~~ +\left[1+ z^{(0)}\right]^2\left[ \frac13 \Theta^{(1)} + \sigma_{\mu\nu}^{(1)}e^\mu_{(0)} e^\nu_{(0)}\right], \\
  && \hat\sigma^{(1)\prime}  +2\hat\theta^{(0)} \hat\sigma^{(1)} = \Psi^{(1)}_0,\label{fx5}\\
  &&\delta_L = \delta_A + 2\frac{z^{(1)}}{1+z^{(0)}},\label{fx6}
\end{eqnarray}
where primes are $\partial/\partial v$.

By (\ref{vofz}) with $H_\|=H$,  (\ref{fx2}) reduces to
the definition $H=\dot a/a$, and then (\ref{fx1})
can be integrated to give (\ref{e-da0}).
Then  (\ref{fx3}) is similar to (\ref{e.66}), with the source term  given by
\begin{equation}
\Phi_{00}^{(1)} = -4\pi G \rho^{(0)}\left[1+ z^{(0)}\right]^2
 \left[\frac{\rho^{(1)}}{\rho^{(0)}} + 2\frac{z^{(1)}}{1+z^{(0)}} \right].
\end{equation}
Since $\hat\theta^{(0)}$ is known,
(\ref{fx5}) can be integrated
once the source (which depends on gradients of the gravitational potentials) is known.

These two derivations have to give the same answer, which they actually do
if the effect of perturbations is not neglected in any of the equations,
and in particular in the equation for $v(z)$.

\subsection{Dyer--Roeder approximation}

It is important to stress that the perturbative description still assumes that the distribution of
matter is continuous (i.e. it assumes that the fluid approximation holds on the
scale of interest),
so that, as long as one is in the weak lensing regime, the whole
effect arises from Ricci focusing with the density of matter equal
to the average cosmic density (because the effect of the shear
appears only at second order).
Zel'dovich~\cite{zel} pointed out that light is actually more
likely to propagate in underdense regions so that an overall
demagnification was expected.
This idea was followed up by \cite{dash,Bertotti,gunn,kant,refs}. The approach came to be named after the later work of Dyer and Roeder~\cite{DR72,DR73,DR74}.

The main assumptions of the DR approximation are: (1)~the Sachs equation~(\ref{s2}) holds; (2)~the relation
$v(z)$ is the FL one,
(\ref{vofz}) with $H_\|=H$; (3)~the null shear $\hat\sigma$ vanishes, as in a
FL universe; (4)~$\Phi_{00}$ is replaced by
$\alpha(z)\Phi_{00}$, where $\alpha(z)$ represents the fraction of (the mean)
matter intercepted by the geodesic bundle. In summary, the DR model assumes
that the bundle is propagating in a FL universe
but that the Ricci focusing is reduced (to reproduce the fact that the beam propagates
mostly in vacuum or underdense regions) and the Weyl focusing remains zero. This implies
that the DR equation is
\begin{eqnarray}
 \frac{\dd^2\bar D_A}{\dd z^2} &+& \left(\frac{\dd\ln H}{\dd z}
 +\frac{2}{1+z}\right) \frac{\dd\bar D_A}{\dd z}
 \nonumber \\
 &=&-\frac{3}{2}\Omega_{\rm m0} 
 \frac{H_0^2}{H^2}(1+z)
 \alpha(z)\bar D_A.
\label{dr-bgd0}
\end{eqnarray}
This attempts to model the global effect of inhomogeneity in a ``mean way'', while
still assuming that the universe is isotropic and homogeneous.
The consistency of the DR approximations has been questioned by Ehlers and Schneider \cite{ehlerss} and others (e.g. \cite{sasaki,Tomita:1999tg,Rasanen:2008be}), independently of Weinberg's photon flux conservation argument \cite{weinberg}.

The smoothness parameter $\alpha$ was initially assumed to be
constant~\cite{DR72,DR73,DR74}. Later it was refined to take into
account its redshift dependence due to the growth of
structure~\cite{linder88,tomita,Mortsell:2001es} and then related to
the statistical properties of large-scale
structure~\cite{Rasanen:2008be,Bolejko:2010nh}. A novel use of the DR approach was proposed in~\cite{Wang:1998eh,Wang:1999bz} to correct SNIa observations for the matter distribution along the line of sight, which has important implications for parameter estimation~\cite{Wang:2011sb}.

On large scales (Mpc) we expect
a distribution of 3D compensated voids giving partially 1D
compensated matter distributions along the line of sight \cite{Bolejko:2010nh}. The lensing
effects will not be the same as a smooth distribution of matter. On
smaller scales (2 kpc to 1 A.U.) we expect to mainly move through
voids,
contaminated by remnant baryonic gas and non-baryonic
dark matter, plus the nearly-smoothly distributed photons and neutrinos. This suggests that the effect can be significant if
matter is clustered on small scales with most of the light beams
used in SNIa observations preferentially moving through
voids.

\subsection{Modifying the DR approximation}

The previous discussions make it explicit that the DR approximation is
not a satisfactory model for the effects of clumping on ray bundles.
Consider the $D_A(z)$ equation, in the absence of pressure and null shear:
\begin{eqnarray}
H_\parallel(z)\frac{\dd}{\dd z}\left[(1+z)^2H_\|(z)\frac{\dd}{\dd z}D_A(z)\right]\nonumber\\ = -4\pi G\rho(z) D_A(z),
\end{eqnarray}
where $H_\parallel(z)=\Theta/3+\sigma_{\mu\nu}e^\mu e^\nu$.
The DR approximation assumes that we can model the encounters of photons with inhomogeneous matter, and not a homogeneous background spacetime, by using $\rho(z)$ as the ``true'' density along a ray -- while leaving the rest of the equation as in the smooth background. Another way to think of this is that the relationship between the affine parameter and redshift is held smooth, and inhomogeneities are assumed not to affect the $v(z)$ relation significantly. However, photons only experience the local curvature, shear and expansion and the average FL-behaviour must somehow emerge from integration along the line of sight.

As discussed above, even in the perturbed FL case the relation $v(z)$ fluctuates, and so the DR relation cannot be relied upon as a useful approximation even in that situation. In particular,
$\alpha$ must depend on the line of sight since each bundle experiences a different
matter profile. In a more general sense, the DR approximation does not account for changes in the local expansion rate due to clumping \cite{Rasanen:2008be}, and does not capture the essence of weak lensing unless $\alpha(z)$ is tuned to a specific form, with no apparent physical motivation~\cite{Bolejko:2010nh}.

We do not aim to provide a detailed analysis of the DR model here. Rather we  offer some possible alternatives in order to estimate how significant the effect of clumping could be, as well as to show how difficult it is to model in a simple but reliable way.

\subsubsection{Modified DR}

In a universe with irrotational dust and arbitrary inhomogeneity, the generalized Friedmann equation is \cite{Tsagas:2007yx}
 \be
{1\over3} \Theta^2=8\pi G\rho+\Lambda +{1\over2}\sigma_{\mu\nu} \sigma^{\mu\nu}- \frac{1}{2}\mathcal{R},
 \ee
where $\mathcal{R}$ is the Ricci curvature scalar of the 3-surfaces orthogonal to the matter four-velocity.
By holding $\Theta$ fixed to the FL background value $3H$, the  DR approximation effectively assumes that the variations of $\rho$ on any null geodesic are compensated by corresponding fluctuations in the shear and curvature, which seems unphysical.

We  expect a photon in the real universe to react to the nonlocal part of the gravitational field created by dark matter halos through local curvature fluctuations in addition to the dynamics of the matter in the intervening space. A reasonable alternative to DR, then, is to first write out the $D_A(z)$ equation in a general FL model, using the Friedmann equation to evaluate $\dd H/\dd z$. Substituting for $H(z)$ using  (\ref{e-fleq}) everywhere, we see that $\Omega_{\rm m}$ appears in several places. Then, replacing $\rho_{\rm m}\to\alpha\rho_{\rm m}$ gives a plausible alternative to the usual DR approximation:
\begin{eqnarray}
 \frac{\dd^2\bar D_A}{\dd z^2} &+& \bigg\{
 \frac{(1+z)H_0}{2\tilde H^2}\left[3\alpha(z)\Omega_{\rm m0}(1+z)+\Omega_{K0}\right]
 \\
 &+&\frac{2}{1+z}\bigg\} \frac{\dd\bar D_A}{\dd z}
 =-\frac{3}{2}\Omega_{\rm m0} 
 \frac{H_0^2}{\tilde H^2}(1+z)
 \alpha(z)\bar D_A\,\nonumber\label{dr-modified}
 \end{eqnarray}
  {where}
 \begin{eqnarray}
 \tilde H(z)^2&=&H_0^2\left[\alpha(z)\Omega_{\rm m0}(1+z)^3 + \Omega_{\Lambda0} + \Omega_{K0}(1+z)^2\right]\nonumber\\
\end{eqnarray}
This modified DR equation attempts to take into account some aspects of the change in expansion expected from an inhomogeneous matter distribution. There are clearly a variety of ways to do this (e.g., we have ignored $\alpha_{,z}$ terms which could be important), but we have chosen just one. See~\cite{Mattsson:2007tj} for an alternative approach.

\subsubsection{Shell approximation}

Consider a single line of sight, smoothed over some scale $\lambda$. The density profile along this line of sight
takes some form $\rho_\lambda(z)$.
If we neglect the angular part of the shear $\sigma_{\mu\nu}$, then we can think of the beam as passing through shells of differing density. There exists a spherically symmetric Lema\^\i tre-Tolman-Bondi (LTB) model with non-zero $\Lambda$ that has the same density profile $\rho_\lambda(z)$ along the past lightcone from the centre. The $D_A(z)$ relation in the LTB model viewed from the centre will approximate the $D_A(z)$ relation along the line of sight we are trying to model. Each line of sight would have a different associated LTB model (a mosaic of cones around us, in the language of~\cite{Wang:1998eh}). The utility of this approximation lies in the fact that we can specify a density profile on a surface of constant time, and use the exact LTB solution to evolve the density backwards onto the past lightcone. We can then calculate $D_A(z)$ exactly for that line of sight. Most importantly, this will account for the variable expansion rate along the direction of propagation which also takes into account the radial component of the shear.

In the LTB model, the angular Hubble rate is given by an effective Friedmann equation~\cite{February:2009pv}
\begin{equation}\label{fltb}
{H_\perp^2(t,r) \over H_{\perp0}^2(r)}=\Omega_{\rm m0}(r)\,a_\perp^{-3}+\Omega_{K0}(r)\,a_\perp^{-2}+\Omega_{\Lambda0}(r), \end{equation}
where the angular scale factor $a_\perp(t,r)$ is normalized to unity today, the $\Omega$'s have an arbitrary radial degree of freedom in them, and $H_{\perp0}(r)$ is calculated once the age is set (or the Hubble rate at the centre is chosen). If $\Omega_{\rm m0}$ is chosen as a constant we have an FL model. To model a radial line of sight we can choose the density profile today as $\rho_0(r)=[1+\delta(r)]\rho_\lambda(0)$, and then
 \be
\Omega_{\rm m0}(r)=\frac{1}{H_{\perp0}^2(r)}\int_0^r \dd r' {r'}^2\rho_0(r')\,.
 \ee
Then (\ref{fltb}) evolves the density back onto the past lightcone, using
\begin{eqnarray}
\label{dtdz}
\frac{\dd t}{\dd z} = -\frac{1}{(1+z)H_\parallel}, ~~~
\frac{\dd r}{\dd z} = \frac{\sqrt{1+\Omega_{K0} H_{\perp0}^2 r^2}}{(1+z)\partial_t\partial_r(a_\perp r)},
\end{eqnarray}
where the radial Hubble rate is $H_\|(t,r)=[\partial_t\partial_r(a_\perp r)]/[\partial_r (a_\perp r)]$. The area distance is then
 \be
D_A(z)=a_\perp(t(z),r(z)) r(z).
 \ee

\subsubsection{Numerical investigation}

We can compare these different approximations numerically. First consider the case of a single density fluctuation.
\begin{figure}[htbp]
\begin{center}
\includegraphics[width=\columnwidth]{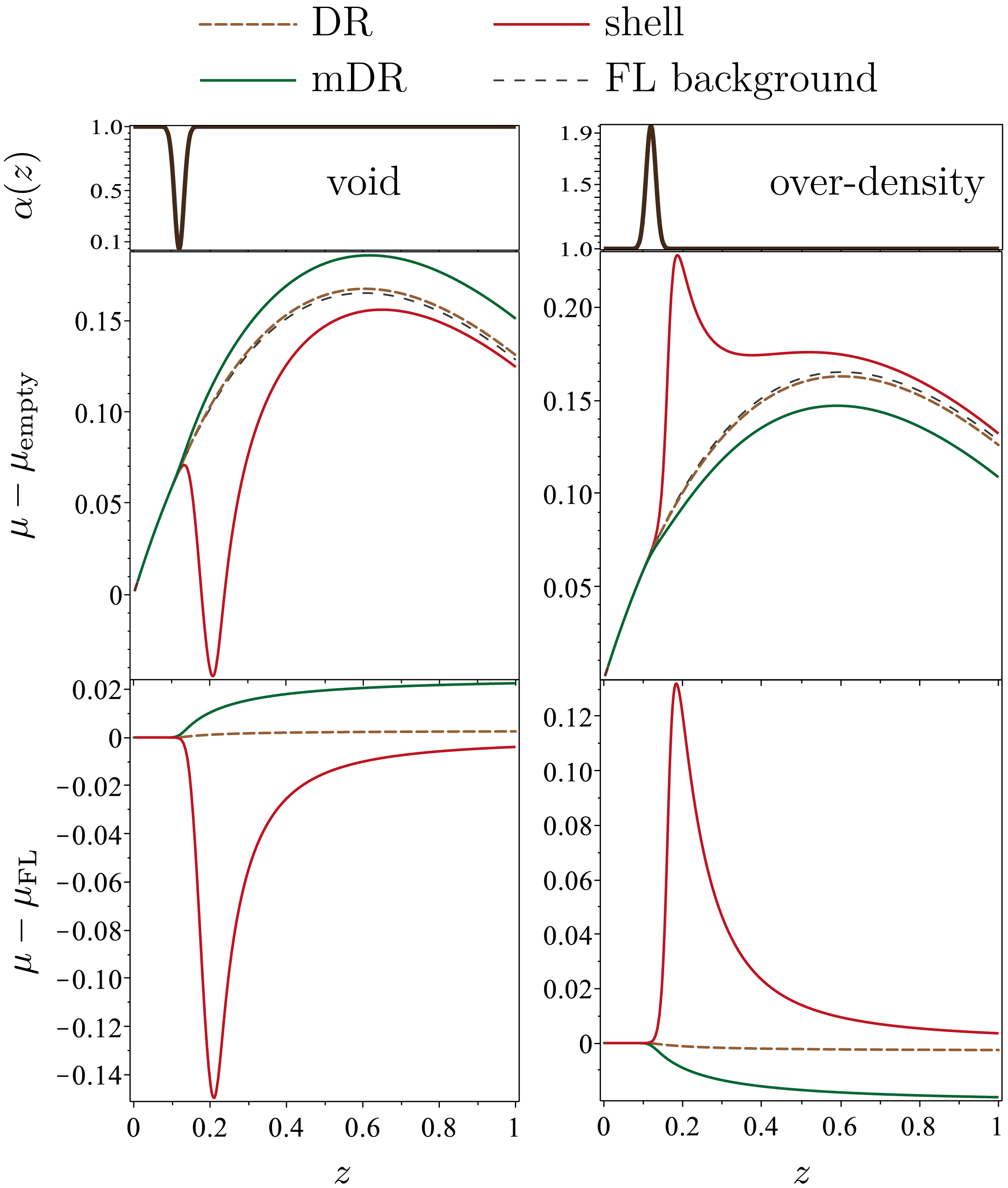}
\caption{The effect of a single void (left) or overdensity (right) on the distance modulus $\mu=5 \log_{10}(D_L/10\,{\rm pc})$ according to three different approximations, using as a `background' a flat LCDM model with $\Omega_m=0.25,~h=0.7$. }
\label{zdr_bump}
\end{center}
\end{figure}
Figure~\ref{zdr_bump} shows the results of looking through a large void and large overdensity, 500\,Mpc away, modelled with a Gaussian deviation from $\alpha=1$ of width $\sim100\,$Mpc. An underdensity causes an increase in the distance modulus at redshifts beyond itself, in both the DR and modified DR cases; the opposite happens for an overdensity.

The DR and modified DR are qualitatively similar, while the shell approximation is very different.
According to the (modified) DR approximations, we should expect SNIa to appear dimmer when located behind an underdensity as compared to an overdensity. By contrast, the shell approximation gives the opposite effect with a much larger amplitude: SNIa located behind a void appear \emph{brighter} than in the fiducial cosmology; located behind an overdensity, they appear dimmer.  The reason is as follows: although a void results in a negatively curved region (which would imply diverging light rays and larger distances), this is accompanied by an increase in the {\em expansion rate} in that region, which actually has a much stronger effect on distances (compare \cite{Rasanen:2008be}). In FL,  increasing the expansion rate and decreasing the density while keeping the age fixed results in a model with smaller distances, and this is exactly what happens here. For an overdensity, the reverse applies. However, note that while the shell approximation captures the mean expansion rate down the line of sight nicely, it may not capture the radial expansion rate correctly. E.g., looking through a spherical void vs a shell with the same density profile have different shear along the line of sight, making our approximation somewhat exaggerated in such a case. 

\begin{figure}[t]
\includegraphics[width=0.40\textwidth]{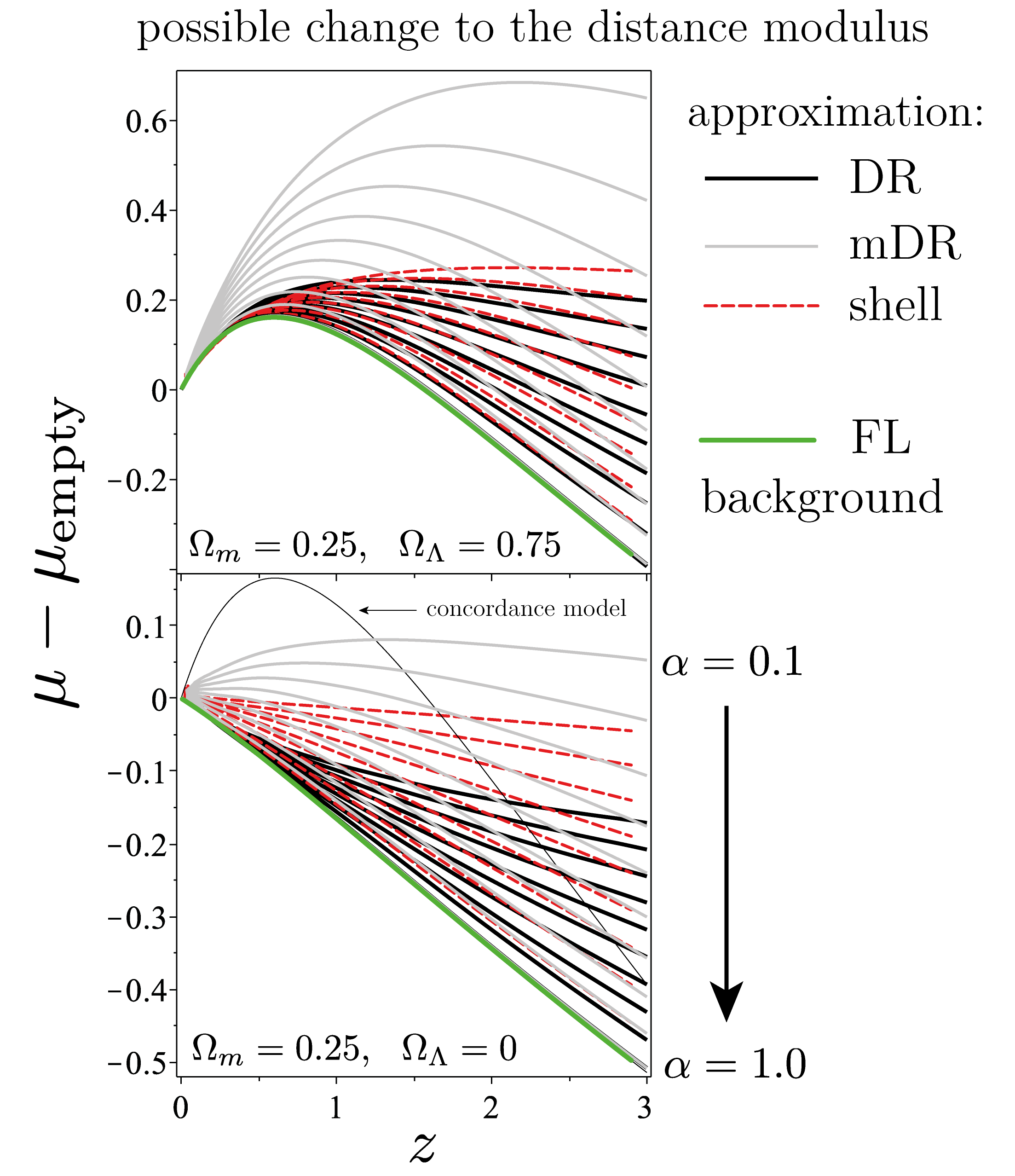}
\caption{
  \label{fig:story}  
  \normalsize  We show the effect of differing
  mean values of $\alpha$  based on different approximations described
  in the text.  The best known of these is  the Dyer-Roeder (DR) which
  simply reduces the energy density in the light propagation equations
  while keeping the background  expansion rate. Attempts to model this
  more  accurately  by  accounting  for  the  varying  expansion  rate
  encountered  along  the  beam  yield  very  different  contributions
  depending on how this is  modelled. One such modified DR (mDR) gives
  dark energy-like behaviour in the  distance modulus even for a model
  with  no  cosmological constant  (all  shown  compared  to an  empty
  model).  
}
\end{figure}

Now consider the case where the density along the line of sight is
reduced by a fixed percentage below the background
value. Figure~\ref{fig:MultiDark} shows that the main contribution of
smoothing a simulation over smaller beam sizes is to reduce the mean
value of $\alpha$, since particles of significant size are rarely
encountered. In the DR and modified DR approximations this amounts to fixing
$\alpha=\,$const. and Monte Carlo-ing over the PDF. In the shell approximation we have $\delta=\alpha-1$
(since $\rho$ is constant, this is really just an FL model with
adjusted parameters). In Fig.~\ref{fig:story} we show the distance
modulus for two fiducial backgrounds: one a standard LCDM model, the
other a curved CDM model with $\Lambda=0$.  We see that all the
approximations give a systematic dimming effect to varying
degrees. (There is actually an overall brightening in the shell case
due to the change in $h$, but we have subtracted this off because SNIa
observations are only sensitive to relative magnitudes, not absolute
ones, which is marginalised over together with $h$.) While the DR and
shell approximations are rather similar, giving changes to the
distance modulus of at most 0.1 at $z\sim 1$, we see that the modified
DR approximation gives a much stronger effect, of several times that. 

\begin{figure}[htbp]
\begin{center}
\includegraphics[width=\columnwidth]{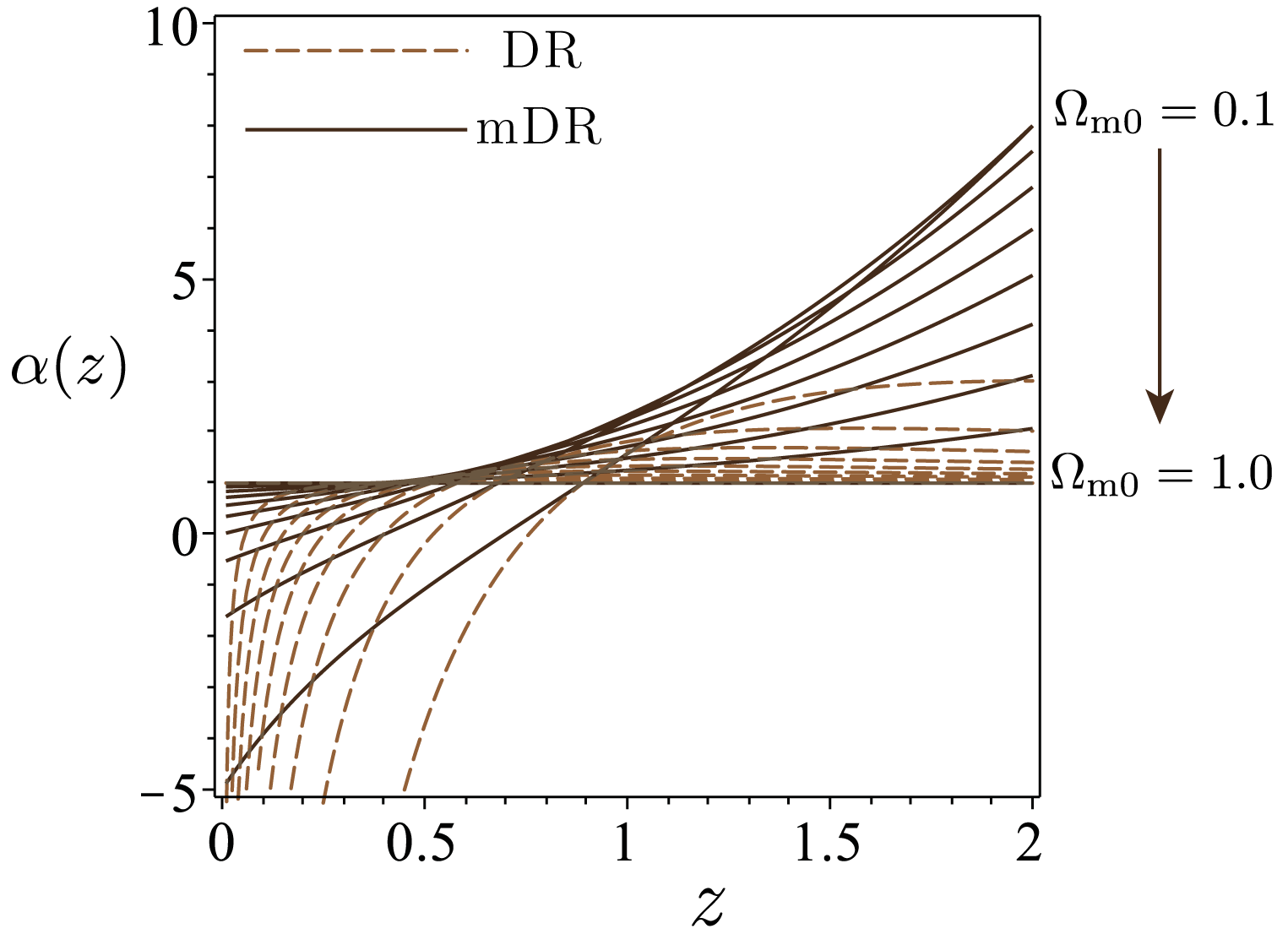}
\caption{The $\alpha(z)$ required to give a LCDM $D_A(z)$ curve, for a variety of $\Omega_{\rm m0}$ (with curvature making up the rest of the energy component). The DR approximation requires extremely negative values of $\alpha$ to mimic dark energy, which might suggest that the effect of modelling narrow beams may not be important for dark energy, and certainly could not be the underlying cause of it. On the other hand, a simple modification to DR yields an approximation that much more readily produces dark energy-like effects. This suggests instead that modelling narrow beams properly, and getting the DR approximation right, could be vital for determining the nature of dark energy.}
\label{fake-lcdm}
\end{center}
\end{figure}
It is striking that in the case $\Lambda=0$ for $\alpha\lesssim0.3$ we see the modified DR distance modulus mimicking the behaviour of a dark energy component. More specifically, both the DR and modified DR approximations can mimic a LCDM model, though the DR approximation requires a drastic change to $\alpha$ at low redshift to do so. For the modified DR approximation this is not so~-- see Fig.~(\ref{fake-lcdm})~-- and it is well known that an LTB distance modulus can mimic any FL one.

It is clear from Fig. \ref{zdr_bump} that the DR
approximation and plausible variations of it can give very different
results -- so that the basis of the DR approximation itself is
suspect. 

\subsection{Other approaches}

Other approaches to the problem of light propagation in a clumpy universe are based on exact nonlinear solutions or numerical methods or both.

In ``Swiss cheese" models, an FL universe contains one or more spherical inhomogeneous regions, with the same average density as the FL model, which could be Schwarzschild vacuoles, or more realistically, LTB balls. These have been used extensively to model the effect of inhomogeneities on light rays (e.g.
\cite{Brouzakis:2006dj,Biswas:2007gi,Marra:2007pm,Vanderveld:2008vi,
Clifton:2009nv,Valkenburg:2009iw,Kantowski:2009jt,Bolejko:2010eb,Szybka:2010ky}).
The results depend on the position and nature of the inhomogeneous regions, as well as on the method for randomization of light rays.
A careful analysis that approximates observations in all directions and over a range of distances \cite{Szybka:2010ky}
concludes that there are only small corrections to the standard FL results. This is not surprising, since the average density along typical lines of sight is very close to the FL density. For the case of SNIa beams, which may preferentially sample underdense regions, the results could be different. However, an inherent limitation of this approach is that, by construction, and given the highly symmetric nature of the inhomogeneities, the clumps do not affect the expansion rate of the universe. 

The method of ray-tracing through N-body simulations (e.g. \cite{Takahashi:2011qd}) is very useful for statistical analysis and predictions for lensing observations. 
However, the dynamical range of scales of matter
inhomogeneities that the simulations can reproduce is very limited.
Ray-tracing within N-body simulations is usually done by projecting all matter
onto equally spaced `lens planes' which are separated
by  about 100 Mpc. Such an approach is not able
to clarify the effect that halos below the simulations mass resolution have on light bundles from SNIa.

Exact solutions that are more general than the Swiss cheese models are usually less realistic, given the highly nonlinear nature of GR. A class of Szekeres models, in which inhomogeneities may be modelled as nonlinear perturbations on an FL background, has been used to investigate light propagation by \cite{Meures:2011gp}.  Despite the idealized nature of clumping, the results show that for inhomogeneities of large spatial size, parameter estimation could be seriously affected (see also~\cite{Krasinski:2010rc}). A stronger conclusion follows from analyzing light rays in a universe with regularly spaced point masses separated by vacuum \cite{Clifton:2009jw}. The average dynamics is close to LCDM, but the optics behaves very differently. Unlike N-body simulations, this model is self-consistent (i.e. the particles generate their own spacetime geometry). However the modelling of matter is necessarily over-simplified.

\section{Ricci and Weyl focusing}

The DR approximation neglects the effect of point sources and
the Weyl focusing they produce, in particular in the strong lensing
regime where it is not negligible. This issue was addressed in
\cite{weinberg}, by considering the effect of the DR equation
(for almost all directions, where Weyl effects can be neglected) and
the effect of strong lensing (for the relatively few directions
where lightrays pass close to matter, with strong lensing occurring
and leading to multiple images). These combined effects are shown to
lead to the usual FL relations when averaged over the whole sky,
the decreased flux in most directions being compensated by higher
flux in a few directions.

Keeping in mind
that strong lensing effects are negligible in most directions, and
in particular for most SNIa (unless we observe a galaxy or a cluster
on the line of sight), we can try to go a step further than the DR
approximation by relating, at least heuristically, the Weyl focusing
to an effective Ricci focusing. The approximation that
the SNIa bundles remain in the weak-field regime can be supported as
follows. If matter is modelled as a gas of particles of 
mass $m$ and proper radius $r_*$, with mean number density $n$, then
the mean energy density is $\rho = m n$. The probability that a
line of sight intersects such a matter particle within the redshift
band $z \to z+\dd z$ is proportional to the surface area of the
particles, to their density, and to the distance light propagates in
this redshift interval, i.e.
\begin{equation}
\dd P = \pi r_*^2 n(z)
\frac{\dd z}{(1+z)H(z)}.
\end{equation}
The average number of particle intersections before the redshift $z$ is the optical depth,
\begin{equation}
\tau(z) = \pi r_*^2
\int_0^z \frac{n(z')}{(1+z')H(z')}\dd z'.
\end{equation}
We assume particle number conservation, $n(z) = n_0(1 + z)^3$, with
$n_0\sim 0.005h^3 {\rm Mpc}^{-3}$ (number density of halos above
$10^{12}\hMsol$ which comprise about 50\% of the mass in the universe) and $r_*\sim   
20h^{-1} {\rm kpc}$ for the central, high density, region of the halos
where the galaxies are assumed to reside. {This implies that at $z=1$, $\tau\sim 0.023$ and $\tau=0.032$, which
means that only 2.3\% and 3.2\% of the lines of sight intersect a galaxy before
$z=1$, respectively for an Einstein-de Sitter and a flat $\Lambda$CDM model with $\Omega_{\rm m}=0.3$. This is a rough estimate -- $\Lambda$ and
inhomogeneous distribution of the luminous matter
would tend to lower it.}

In order to describe the transition from Weyl to Ricci focusing, we
first recall the lensing effect of a single mass, whose  gravitational potential is
 \be
\Phi =-Gm\big({b^2+u^2}\big)^{-1/2},
 \ee
where $b$ is the
impact parameter and $u$ the distance along the line of sight. The
deflection angle is thus
\begin{equation}
{\bm \alpha} =4Gm\frac{\bm b}{b^2}.
\end{equation}
The critical density and the Einstein angular radius are
respectively defined by $\Sigma_{\rm crit}=
{D_{OS}}/{(4\pi G D_{OL}D_{LS})}$ and $\theta_{\rm E}^2=4Gm
{D_{LS}}/{(D_{OL}D_{OS})}$, so that the lens equation takes the
form
\begin{equation}
{\bm\theta}={\bm\theta}_{\rm s} + \hat{\bm\alpha}({\bm\theta}),
\end{equation}
where $D_{LS}$, $D_{OL}$ and $D_{OS}$ are the angular distances
between the lens plane and the source, the observer and the lens
plane, and the observer and the source. The angular position of the
image on the lens plane is then given by ${\bm\theta}={\bm
b}/D_{OL}$. $\hat{\bm\alpha}$ is given by $\hat{\bm\alpha}= {\bm\theta}{\theta_{\rm E}^2}/{\theta^2}$, since
$\Sigma({\bm\theta})=m\delta^{(2)}({\bm b})=m\delta^{(2)}({\bm
\theta})/D_{OL}^2$ for a point mass. The amplification matrix is
then obtained as ${\cal A}^a_b=\partial\theta_{\rm
s}^a/\partial\theta^b$, so that ${\cal A}_{ab}=\delta_{ab}
-\partial_{\theta^a}\hat\alpha_b=
 \delta_{ab} -\partial_{\theta^a}\partial_{\theta^b}\psi$
with
\begin{equation}
\psi=\theta_{\rm E}^2\ln\theta.
\end{equation}

We conclude that the convergence and the shear are given by
\begin{equation}
 \kappa =\pi \theta_{\rm E}^2 \delta^{(2)}({\bm\theta}),\quad ~
 (\gamma_1,\gamma_2)=\frac{\theta_{\rm E}^2}{\theta^4}
(\theta_2^2-\theta_1^2,2\theta_1\theta_2),
\end{equation}
which is more easily written in terms of the polar angle on the
screen ($\theta_1=\theta\cos\varphi$, $\theta_2=\theta\sin\varphi$)
as
\begin{equation}
(\gamma_1,\gamma_2)=-\frac{\theta_{\rm E}^2}{\theta^2}
(\cos2\varphi ,\sin2\varphi ).
\end{equation}

For a single mass and a line of sight that does not
intersect it, $\kappa = 0$ and the magnification is
$\mu =[1-\vert\gamma\vert^2]^{-1}\sim 1+ 2 {\theta_{\rm E}^4}{\theta^{-4}}$,
neglecting the image inside the Einstein radius.
Now consider a shell of thickness $\dd z$ such that the density of
particles is $n(z)$ and such that $\dd\chi\ll D_{LS},
D_{OL},D_{OS}$. For a typical lightray the total amplification
matrix will have a shear given by
\begin{equation}
 \big[\hat\gamma_1(z),\,\hat\gamma_2(z) \big]= \theta_{\rm E}^2\sum\Big[ \frac{\cos2\varphi}{\theta^2},\,\frac{\sin2\varphi}{\theta^2}\Big],
\end{equation}
where the sum is over all particles of the shell. If the
particles are distributed homogeneously, this implies that they are
isotropically distributed around the line of sight, so that we expect $\hat{\gamma}_1(z), \hat{\gamma}_2(z)\sim 0$.

Intuitively, this is understood by the fact that each point mass
induces an ellipticity in a different direction and they
should average to reduce the total ellipticity. The remaining effect
of all the Weyl distortions is thus an effective convergence that
can be determined from the magnification $\hat{\kappa}(z) = \theta_{\rm E}^4\sum{\theta^{-4}}$.
Estimating the sum by assuming we have a uniform distribution and
that the typical smallest distance is $\theta\sim
n^{-1/3}/D_{OL}$, we get that $\sum{\theta}^{-4}\sim D_{OL}^3n$,
and thus
 \be
\hat{\kappa}(z)\sim f_K^3[\chi(z)] n(z)\left\lbrace4G m
\frac{f_K[\chi(z_{\rm s})-\chi(z)]}{f_K[\chi(z)]f_K[\chi(z_{\rm
s})]} \right\rbrace^2\,,
 \ee
 where $f_K(\chi)$ is the unfilled angular
distance, i.e. using $\hat H$
instead of $H$ in  (\ref{chiz}).
We can compare to the effect that a homogeneous distribution of density
$n_{\rm eff}(z)=\alpha(z)n(z)$ would generate through its Ricci
focusing to get
\begin{equation}
 \alpha(z)= 4Gm\frac{f_K[\chi(z_{\rm s})-\chi(z)]}{f_K[\chi(z_{\rm s}]}H(z).
\end{equation}
This is the effective DR parameter for the averaged Ricci
convergence. It stills depend explicitly on $m$ because this is a second order effect, hence scaling as $\theta_{\rm E}^4$, while
it is first order in a homogeneous medium. Such an estimate is indeed very crude but it confirms the statements
of  \cite{DR81,weinberg} and the results of the numerical
simulations of  \cite{HW97}.

\section{Discussion and conclusions}

The effect of the inhomogeneity of the matter distribution
induces a dispersion of the magnification of SNIa and thus of
the luminosity distance. We have argued that this effect has not been
properly modelled for SNIa since the beam is very narrow and far
below the scales resolved in any numerical simulation. 

For the first time, we have attempted to quantify the probability distribution for narrow beams, using a combination of N-body simulations and a PS approach. For a narrow beam of fixed length, the PDF is non-Gaussian, peaked at densities below the cosmic mean, with a power-law tail, whose power depends on the diameter of the beam, describing the relatively few lines of sight which have an over-dense mean. These PDFs contrast sharply with distributions based on using cubes of the same volume. These estimations are based on current N-body simulations which do not have the resolution to probe beams with a diameter $\lesssim100\,h^{-1}$\,kpc. Nevertheless, the trend is clear: narrow beams typically experience a lower than average density, and do not sample the cosmic mean density until their length approaches the Hubble scale. Based on our results, we estimate that significantly more than 75\% of beams experience less than the mean density.

From a theoretical point of view, the effect must be described by the set
of equations~(\ref{e.1}), (\ref{s2}) and~(\ref{vofz}) that describe the distortion
and magnification of any light bundle, whatever the spacetime geometry. The explicit covariant form of these equations is given by (\ref{zsig})--(\ref{zda2}).
The main problem is that the solution of this set of equations requires
a description of the distribution of matter on the scale of the beam size, i.e. on scales much smaller than those of our current understanding.

This dispersion has been modelled in some regimes but we argue that the
distribution of matter on the scales relevant for the description of
a SNIa bundle is not understood yet. On such small scales, the
statistical dispersion comes with a bias that has two origins: (1)~the fact that the nonlinear terms in the expression for
the magnification can  not be neglected a priori; (2)~an
observational selection effect due to the fact that most SNIa
are observed in directions where they are not overshadowed by a
galaxy. The bundles included in SNIa analysis are thus more
likely to probe under-dense regions.

Why is the Hubble diagram so compatible with that
of a Friedmann-Lema\^{\i}tre universe?
In particular,  the typical transverse size of the
bundle is smaller than the typical mean distance between the smallest
bound structures in the currently favoured CDM paradigm.
Why does the fluid approximation used to interpret
the data hold? This question is two-sided. It questions the robustness of the
interpretation of the cosmological data but also offers a way to constrain
the distribution of matter on small scales. We also gave a heuristic argument
concerning the Ricci-Weyl focusing issue, leading to a prediction of an effective DR factor $\alpha(z)$ once the fractions of clustered  and
smooth matter are known.

Another description is provided by the Dyer-Roeder equation. It has however
some simplifying hypotheses that neglect the effects of changes in $v(z)$ and the local expansion due to clumping. We suggested two plausible modifications to the DR approach, and showed that the 3 models produce very different results~-- thus undermining confidence in the DR approximation. In particular, it is not clear whether under-densities lead to demagnification (due to negative curvature) or magnification (due to the increase in the expansion rate). Our shell approximation clearly points to the opposite effect calculated via normal lensing or the DR approximation: a SN located just behind an under-density should appear brighter than it would in the fiducial cosmology. In fact, we estimate that a SN the far side of a 100\,Mpc void could be 0.1\,mag brighter than it would be with not void present. Though likely an over-estimation, this could have important implications for parameter estimation from SNIa. Quantifying this properly is an important open problem. In fact, it is striking to note from our investigations of N-body simulations that we should expect $\alpha$ to vary as a function of radius from us from significantly below unity locally, approaching unity only on Hubble scales. Within our shell approximation, this would give exactly the kind of model used in Hubble scale `void models', which require no dark energy, but without any kind of anti-Copernican fine tuning involved~\cite{2012arXiv1204.5505C}. Every observer would observe such an effect.

While accurate modelling of such beams may be problematic for some time, we can still observationally test to see if there are problems and phenomenologically correct for them. 
\begin{description}
\item[SNIa line of sight] Dividing up SNIa samples according to the estimated density along the line of sight may reveal a bias. If so, this may indicate that the effects we have discussed here must be taken into account. 

\item[discordant distances]  In any exact relativistic model the luminosity distance is $D_L=(1+z)^2D_A$, where $D_A$ is the  area angular diameter distance. Large scale measurements of the area distance will not be affected by the problems we have discussed here, however, so we can expect a failure at some level of this relation when comparing measurements of $D_A$ from large scale measurements such as the BAO and the CMB to measurements of $D_L$ from SNIa. On smaller scales, where the area distance is measured from radio or quasar sources, there could be an effective reciprocity breakdown because such sources still have much larger beam sizes than SNIa, and so smear the matter distribution to include many more over-dense regions. Such a violation was found in~\cite{Bassett:2003vu}, where a relative brightening of SNIa was found~-- as we would predict from the shell approximation.  

\item[consistency conditions] A variety of consistency tests have recently been developed as a way of testing the standard model (see~\cite{2012arXiv1204.5505C} for a review).  It has recently been shown that these are strongly sensitive to changes of the DR form~\cite{2012arXiv1204.1083B}. This implies that they can be used to probe the effects we have discussed here. 
\end{description}
Generically, then, distances to the same object will depend on the scale over which the light from the source smears the intervening matter distribution.

We have found that the old problem of modelling narrow beams remains unsolved. As different interpretations of the problem give conflicting yet significant effects, we believe this problem needs considerably more attention. This is important not only from a theoretical perspective, but to ensure precision cosmology delivers correct answers as well as precise ones.

~\\{\bf Acknowledgements:}\\
We thank David Bacon, Krzysztof Bolejko, Peter Coles, Ruth Durrer, Stefan Hilbert, Yannick Mellier, Bob Nichol, Brian Schmidt, Joe Silk, Zachary Slepian and Yun Wang for discussions and/or comments. RM is supported by a South African SKA Research Chair, and by the UK Science \& Technology Facilities Council (grant no. ST/H002774/1). CC, GE, AF, RM, OU are supported by a NRF (South Africa)/ Royal Society (UK) exchange grant. JPU is partially supported by the ANR-Thales. The Millennium Simulation  and the MultiDark databases used in this paper and  the web  application providing online  access to  them were constructed  as part  of the  activities of  the  German Astrophysical Virtual  Observatory.   Data  for  halos  and  galaxies  are  publicly available at http://www.mpa-garching.mpg.de/millennium and http://www.multidark.org/MultiDark. The MultiDark database is the result of a collaboration between the Leibniz-Institute for Astrophysics Potsdam (AIP) and the Spanish MultiDark Consolider Project CSD2009-00064. The Bolshoi and MultiDark simulations were run on the NASA's Pleiades supercomputer at the NASA Ames Research Center.


\begin{references}

\bibitem{SNIa}
  R.~Amanullah {\it et al.},
  ``Spectra and Light Curves of Six Type Ia Supernovae at $0.511 < z < 1.12$ and
  the Union2 Compilation,''
  Astrophys.\ J.\  {\bf 716}, 712 (2010)
  [arXiv:1004.1711].


\bibitem{zel}
Ya. B. Zel'dovich, ``Observations in a universe homogeneous in the mean", Soviet Astr. AJ {\bf 8}, 13 (1964).

\bibitem{feyn}
R. P. Feynman, unpublished talk (1964). (See \cite{gunn}.)

\bibitem{dash}
V. M. Dashevskii and V. I. Slysh,
``On the Propagation of Light in a Nonhomogeneous Universe",
Soviet Astr. AJ {\bf 9}, 671 (1966).

\bibitem{Bertotti}
  B. Bertotti,
  ``The luminosity of distant galaxies,''
 Proc. Roy. Soc. (London) A {\bf 294}, 195 (1966).

\bibitem{gunn}
J. E. Gunn, ``A fundamental limitation on the accuracy of angular measurements in observational cosmology", Astrophys. J. {\bf 147}, 61 (1967); J. E. Gunn, ``On the propagation of light in inhomogeneous cosmologies. I. Mean effects", Astrophys. J. {\bf 150}, 737 (1967).

\bibitem{kant}
R. Kantowski,
``Corrections in the luminosity-redshift relations of the homogeneous Friedmann models",
Astrophys. J. {\bf 155}, 89 (1969).

\bibitem{refs}
S. Refsdal,
``On the propagation of light in universes with inhomogeneous mass distribution",
Astrophys. J. {\bf 159}, 357 (1970).

\bibitem{kantowski}
  R.~Kantowski, T.~Vaughan and D.~Branch,
  ``The Effects of Inhomogeneities on Evaluating the Deceleration Parameter q$_0$,''
  Astrophys.\ J.\  {\bf 447}, 35 (1995)
  \url{arXiv:astro-ph/9511108}.

\bibitem{jf97}
  J.A. Frieman,
  ``Weak Lensing and the Measurement of q0 from Type Ia Supernovae'',
  Comm. Astrophys. {\bf 18}, 323 (1998)
  \url{astro-ph/9608068}.

\bibitem{Wang:1998eh}
  Y.~Wang,
  ``Supernova pencil beam survey,''
  Astrophys.\ J.\  {\bf 531}, 676 (2000)
  [astro-ph/9806185].

\bibitem{Wang:1999bz}
  Y.~Wang,
  ``Flux-averaging analysis of type ia supernova data,''
  Astrophys.\ J.\  {\bf 536}, 531 (2000)
  [astro-ph/9907405].

\bibitem{Wang:2004ax}
  Y.~Wang,
  ``Evidence for weak lensing of supernovae,''
  JCAP {\bf 0503}, 005 (2005)
  [astro-ph/0406635].

\bibitem{hl}
  D.~E.~Holz and E.~V.~Linder,
 ``Safety in numbers: Gravitational Lensing Degradation of the Luminosity
  Distance-Redshift Relation,''
  Astrophys.\ J.\  {\bf 631}, 678 (2005)
  \url{arXiv:astro-ph/0412173}.

\bibitem{Cooray:2005yp}
  A.~Cooray, D.~Holz and D.~Huterer,
  ``Cosmology from supernova magnification maps,''
  Astrophys.\ J.\  {\bf 637}, L77 (2006)
  \url{arXiv:astro-ph/0509579}.

\bibitem{Dodelson:2005zt}
  S.~Dodelson and A.~Vallinotto,
  ``Learning from the Scatter in Type Ia Supernovae,''
  Phys.\ Rev.\  D {\bf 74}, 063515 (2006)
  \url{arXiv:astro-ph/0511086}.

\bibitem{Shapiro:2009sr}
  C.~Shapiro, D.~Bacon, M.~Hendry and B.~Hoyle,
  ``Delensing Gravitational Wave Standard Sirens with Shear and Flexion Maps,''
  Mon. Not. Roy. Astron. Soc. {\bf 404}, 858 (2010)
  [arXiv:0907.3635];
  C.~M.~Hirata, D.~E.~Holz and C.~Cutler,
  ``Reducing the weak lensing noise for the gravitational wave Hubble diagram using the non-Gaussianity of the magnification distribution,''
  Phys.\ Rev.\  {\bf D81}, 124046 (2010)
  [arXiv:1004.3988].
  
\bibitem{Dalal2002}
  N.~Dalal, D.~E.~Holz, X.~l.~Chen and J.~A.~Frieman,
  ``Corrective lenses for high redshift Supernovae,''
  Astrophys.\ J.\  {\bf 585}, L11 (2003)
  \url{arXiv:astro-ph/0206339}.


\bibitem{fb97}
 F. Bernardeau,
 ``Weak lensing detection in CMB maps'',
 Astron. Astrophys. {\bf324}, 15 (1997)

\bibitem{Mellier:1998pk}
  Y.~Mellier,
  ``Probing the Universe with Weak Lensing,''
  Ann.\ Rev.\ Astron.\ Astrophys.\  {\bf 37} (1999) 127,
  \url{arXiv:astro-ph/9812172}.

\bibitem{js97}
 B. Jain and U. Seljak,
 ``Cosmological model predictions for weak lensing: Linear and nonlinear regimes'',
 Astrophys. J. {\bf484}, 560 (1997) [arXiv:astro-ph/9611077].

\bibitem{KM:2009}
  K.~Kainulainen and V.~Marra,
  ``SNe observations in a meatball universe with a local void,''
  Phys.\ Rev.\  D {\bf 80}, 127301 (2009)
  [arXiv:0906.3871].

\bibitem{KM:2010}
  K.~Kainulainen and V.~Marra,
  ``Accurate Modeling of Weak Lensing with the sGL Method,''
  Phys.\ Rev.\  D {\bf 83}, 023009 (2011)
  \url{arXiv:1011.0732}.

\bibitem{weinberg}
 S. Weinberg,
 ``Apparent luminosities in a locally inhomogeneous universe,''
   Astrophys.\ J.\  {\bf 208}, L1 (1976).

\bibitem{PressSchechter1974}
  W.~P.~Press and P.~Schechter,
 ``Formation of Galaxies and Clusters of Galaxies by Self-Similar Gravitational Condensation'',
 Astrophys. J. {\bf 187}, 425 (1974).

\bibitem{JenkinsEtAL2001}
  A.~Jenkins, C.~S.~Frenk, S.~D.~M.~White, J.~M.~Colberg, S.~Cole, A.~E.~Evrard, H.~M.~P.~Couchman and N.~Yoshida,
  ``The mass function of dark matter haloes'',
  Mon. Not. Roy. Astron. Soc. {\bf 321}, 372 (2001)
  \url{arXiv:astro-ph/0005260}.

\bibitem{ReedEtAl2003}
  D.~Reed, J.~Gardner, T.~Quinn, J.~Stadel, M.~Fardal, G.~Lake and F.~Governato,
  ``Evolution of the mass function of dark matter haloes'',
  Mon. Not. Roy. Astron. Soc. {\bf 346}, 565 (2003)
  \url{arXiv:astro-ph/0301270}.

\bibitem{WarrenEtAl2006}
  M.~S.~Warren, K.~Abazajian, D.~E.~Holz and L.~Teodoro,
  ``Precision Determination of the Mass Function of Dark Matter Halos'',
   Astrophys. J {\bf 646}, 881 (2006)
   \url{arXiv:astro-ph/0506395}.

\bibitem{ReedEtAl2007}
  D.~S.~Reed, R.~Bower, C.~S.~Frenk, A.~Jenkins and T.~Theuns,
  ``The halo mass function from the dark ages through the present day'',
  Mon. Not. Roy. Astron. Soc. {\bf 374}, 2 (2007)
  \url{arXiv:astro-ph/0607150}.

\bibitem{TinkerEtAl2008}
J.~Tinker, A.~V.~Kravtsov, A.~Klypin, K.~Abazajian, M.~Warren, G.~Yepes, S.~Gottlöber, and D.~E.~Holz,
  ``Toward a Halo Mass Function for Precision Cosmology: The Limits of Universality'',
   Astrophys. J {\bf 688}, 709 (2008)
   \url{arXiv:0803.2706}.

\bibitem{AnderhaldenDeiemnd2011}
  D.~Anderhalden and J.~Diemand,
  ``The total mass of dark matter haloes'',
  Mon. Not. Roy. Astron. Soc. {\bf 414}, 3166 (2011)
  \url{arXiv:1102.5736}.

\bibitem{FaltenbacherEtAl2010}
  A.~Faltenbacher, A.~Finoguenov and N.~Drory,
 ``The Halo Mass Function Conditioned on Density from the Millennium Simulation'',
 Astrophys. J. {\bf 712}, 484 (2010)
 \url{arXiv:1002.0844}.

\bibitem{MoreEtAl2011}
  S.~More,  A.~V.~Kravtsov, N.~Dalal and S.~Gottlöber,
  ``The overdensity and masses of the friends-of-friends halos and universality of the halo mass function'',
  \url{arXiv:1103.0005}.

\bibitem{PradaEtAl2006}
  F.~Prada, A.~Klypin, E.~ Simonneau, J.~ Betancort-Rijo, S.~Patiri, S.~ Gottl{\"o}ber, S. and  M.~A.~Sanchez-Conde,
  ``How Far Do They Go? The Outer Structure of Galactic Dark Matter Halos'',
  Astrophy. J. {\bf 645}, 1001 (2006)
  \url{arXiv:astro-ph/0506432}.

\bibitem{CuestaEtAl2008}
  A.~J.~Cuesta, F.~Prada, A.~Klypin and M.~Moles,
  ``The virialized mass of dark matter haloes'',
  Mon. Not. Roy. Astron. Soc. {\bf 389}, 385 (2008)
  \url{arXiv:0710.5520}


\bibitem{KomatsuEtAl2009}
   E.~Komatsu, J.~Dunkley, M.~R.~Nolta et al. 
   ``Five-Year Wilkinson Microwave Anisotropy Probe Observations:
   Cosmological Interpretation'', 
   ApJS {\bf 180}, 330 (2009)
   \url{arXiv:0803.0547}

\bibitem{Metcalf1999}
  R.~B.~Metcalf and J.~Silk,
  ``A fundamental test of the nature of dark matter,''
  Astrophys.\ J.\  {\bf 519}, L1 (1999)
  \url{arXiv:astro-ph/9901358}.

\bibitem{Metcalf2006}
  R.~B.~Metcalf and J.~Silk,
  ``New Constraints on Macroscopic Compact Objects as a Dark Matter
  Candidate from Gravitational Lensing of Type Ia Supernovae,''
  Phys.\ Rev.\ Lett.\  {\bf 98}, 071302 (2007)
  \url{arXiv:astro-ph/0612253}.

\bibitem{sw87}
 P. Schneider, and R.V. Wagoner,
 ``Amplification and polarization of Supernovae by gravitational lensing'',
 Astrophys. J. {\bf314}, 154 (1987).

\bibitem{rauch}
 K.P. Rauch,
 ``Gravitational microlensing of high-redshift Supernovae by compact objects'',
  Astrophys. J. {\bf374}, 83 (1991).

\bibitem{HW97}
  D.~E.~Holz and R.~M.~Wald,
  ``A New method for determining cumulative gravitational lensing effects in inhomogeneous universes,''
  Phys.\ Rev.\  D {\bf 58}, 063501 (1998)
  \url{arXiv:astro-ph/9708036}.

\bibitem{sh99}
 U. Seljak, and D.E. Holz,
 ``Limits on the density of compact objects from high redshift Supernovae'',
 Astron. Astrophys. Lett. {\bf351}, L10 (1999)
 \url{astro-ph/9910482}.

\bibitem{BertoneEtAl2005}
  G.~Bertone, D.~Hooper and J.~Silk,
  ``Particle dark matter: evidence, candidates and constraints'',
  Phys.\ Rep. {\bf 405}, 279 (2005)
  \url{arXiv:hep-ph/0404175}.

\bibitem{DiemandEtAl2005}
  J.~Diemand, B.~Moore and J.~Stadel,
  ``Earth-mass dark-matter haloes as the first structures in the early Universe'',
  Nature {\bf 433}, 389 (2005)
  \url{arXiv:astro-ph/0501589}.

\bibitem{Zentner2005}
  A.~R.~Zentner,
  ``The Excursion Set Theory of Halo Mass Functions, Halo Clustering, and Halo Growth'',
  Int.\ J.\  Mod.\ Phys.\ D {\bf16} 763 (2007)
  \url{arXiv:astro-ph/0611454}.

\bibitem{AnguloWhite2010}
  R.~E.~Angulo and  S.~D.~M.~White,
  ``The birth and growth of neutralino haloes'',
  Mon. Not. Roy. Astron. Soc. {\bf 401}, 1796 (2010)
  \url{arXiv:0906.1730}.


 \bibitem{Klypin:2011}
   A.~,A.~Klypin, S.~Trujillo-Gomez, S.~Primack, 
   ``Dark Matter Halos in the Standard Cosmological Model: Results from the Bolshoi Simulation''
   ApJ {\bf 740}, 102 (2011) 
   \url{ arXiv:1002.3660}

\bibitem{Springel:2005} 
  V.~Springel, S.~D.~M.~White, A.~Jenkins at al.,  
  ``Simulations of the formation, evolution and clustering of galaxies and quasars''
  Nature {\bf 435}, 629 (2005) 
  \url{arXiv:astro-ph/0504097}

\bibitem{Prada:2011}
  F.~Prada, A.~,A.~Klypin, A.~Cuesta et al., 
  ``Halo concentrations in the standard LCDM cosmology''
  \url{arXiv:1104.5130}

\bibitem{Lemson:2006}
  G.~Lemson \& Virgo Consortium, 
  ``Halo and Galaxy Formation Histories from the Millennium
  Simulation: Public release of a VO-oriented and SQL-queryable
  database for studying the evolution of galaxies in the LambdaCDM
  cosmogony'' 
  \url{arXiv:astro-ph/0608019} 

\bibitem{Riebe:2011}
  K.~Riebe, A.~Partl, H.~Enke et al., 
  ``The MultiDark Database: Release of the Bolshoi and MultiDark Cosmological Simulations''
  \url{arXiv:1109.0003}

\bibitem{Masaki:2011yc}
  S.~Masaki, M.~Fukugita and N.~Yoshida,
  ``Matter Distribution around Galaxies,''
  [arXiv:1105.3005].



\bibitem{sachs}
R. Sachs, Gravitational waves in General Relativity. VI. The outgoing radiation condition", Proc. Roy. Soc. Lond. A {\bf 264}, 309 (1961).

\bibitem{DR72}
  C. C. Dyer, and R. C. Roeder,
  ``The distance-redshift relation for universes with no intergalactic medium,''
  Astrophy. J. {\bf 174}, L115 (1972).

\bibitem{DR73}
  C. C. Dyer, and R. C. Roeder,
  ``Distance-redshift relations for universes with some intergalactic medium,''
  Astrophy. J. {\bf 180}, L31 (1973).

\bibitem{DR74}
  C. C. Dyer, and R. C. Roeder,
  ``Observations in locally inhomogeneous cosmological models,''
  Astrophys. J. {\bf 189}, 167 (1974).

\bibitem{DR81}
  C.C. Dyer, and R.C. Roeder,
  ``On the transition from Weyl to Ricci focusing,''
  Gen. Relat. Grav. {\bf 13}, 1157 (1981).

\bibitem{fangwu}
L. Fang and X. Wu, ``Geometrical optics in an inhomogeneous universe", Chinese. Phys. Lett. {\bf 6}, 223 (1989); X. Wu,
``The magnitude-redshift relation in a locally inhomogeneous universe", Astron. Astrophys. {\bf 239}, 29 (1990).

\bibitem{Ellis:1998ha}
  G.~F.~R.~Ellis, B.~A.~Bassett and P.~K.~S.~Dunsby,
  ``Lensing and caustic effects on cosmological distances,''
  Class.\ Quant.\ Grav.\  {\bf 15}, 2345 (1998)
  [arXiv:gr-qc/9801092].

\bibitem{Rose:2001qi}
  H.~G.~Rose,
  ``Apparent magnitudes in an inhomogeneous universe: The Global viewpoint,''
  Astrophys.\ J.\  {\bf 560}, L15 (2001)
  [arXiv:astro-ph/0106489].

\bibitem{Kibble:2004tm}
  T.~W.~B.~Kibble and R.~Lieu,
  ``Average magnification effect of clumping of matter,''
  Astrophys.\ J.\  {\bf 632}, 718 (2005)
  [arXiv:astro-ph/0412275].

\bibitem{Kostov:2009uc}
  V.~Kostov,
  ``Average luminosity distance in inhomogeneous universes,''
  JCAP {\bf 1004}, 001 (2010)
  [arXiv:0910.2611].

\bibitem{ppu}
 C. Pitrou, T. Pereira, and J.-P. Uzan,
 ``Weak lensing B-modes on all scales as a probe of local isotropy'',
 [arXiv:1204.1203.6029].

\bibitem{Clarkson:2010uz}
  C.~Clarkson and R.~Maartens,
  ``Inhomogeneity and the foundations of concordance cosmology,''
  Class.\ Quant.\ Grav.\  {\bf 27}, 124008 (2010)
  [arXiv:1005.2165].

\bibitem{cdmu}
C. Clarkson, R. Durrer, R. Maartens and O. Umeh, in preparation.

\bibitem{Tsagas:2007yx}
  C.~G.~Tsagas, A.~Challinor and R.~Maartens,
  ``Relativistic cosmology and large-scale structure,''
  Phys.\ Rept.\  {\bf 465}, 61 (2008)
  [arXiv:0705.4397].

\bibitem{sasaki}
M. Sasaki, ``Cosmological gravitational lens equation: its validity and limitation", Prog. Theor. Phys. {\bf 90}, 753 (1993).

\bibitem{Tomita:1999tg}
  K.~Tomita, H.~Asada and T.~Hamana,
  ``Distances in inhomogeneous cosmological models,''
  Prog.\ Theor.\ Phys.\ Suppl.\  {\bf 133}, 155 (1999)
  [arXiv:astro-ph/9904351].

\bibitem{Rasanen:2008be}
  S.~Rasanen,
  ``Light propagation in statistically homogeneous and isotropic dust universes,''
  JCAP {\bf 0902}, 011 (2009)
  [arXiv:0812.2872].

\bibitem{Rasanen:2009uw}
  S.~Rasanen,
  ``Light propagation in statistically homogeneous and isotropic universes with general matter content,''
  JCAP {\bf 1003}, 018 (2010)
  [arXiv:0912.3370].

\bibitem{schneiderbook}
 P. Schneider, J. Ehlers, and E.~E. Falco,
 {\it Gravitational Lenses}
 (Springer, 1992).

\bibitem{pubook}
 P. Peter and J.-P. Uzan, {\it Primordial cosmology} (Oxford Univ. Press, 2009).

\bibitem{Perlick2004}
  V.~Perlick,
  ``Gravitational lensing from a spacetime perspective,''
  Living Rev.\ Rel.\  {\bf 7}, 9 (2004).

\bibitem{Bonvin:2005ps}
  C.~Bonvin, R.~Durrer and M.~A.~Gasparini,
  ``Fluctuations of the luminosity distance,''
  Phys.\ Rev.\  D {\bf 73}, 023523 (2006)
  \url{arXiv:astro-ph/0511183}.

\bibitem{ubm}
J.-P. Uzan, F. Bernardeau, and Y. Mellier,
``Time drift of cosmological redshifts and its variance,''
 Phys. Rev. D {\bf77} 021301(R) (2008)
\url{arXiv:0711.1950}.

\bibitem{Cooray:2005yr}
  A.~Cooray, D.~Huterer and D.~Holz,
  ``Problems with Pencils: Lensing Covariance of Supernova Distance
  Measurements,''
  Phys.\ Rev.\ Lett.\  {\bf 96}, 021301 (2006)
  \url{arXiv:astro-ph/0509581}.

\bibitem{Sarkar:2007sp}
  D.~Sarkar, A.~Amblard, D.~E.~Holz and A.~Cooray,
  ``Lensing and Supernovae: Quantifying The Bias on the Dark Energy Equation of
  State,''
  \url{arXiv:0710.4143}.

\bibitem{Cooray:2008qn}
  A.~R.~Cooray, D.~E.~Holz and R.~Caldwell,
  ``Measuring dark energy spatial inhomogeneity with supernova data,''
  JCAP {\bf 1011}, 015 (2010)
  \url{arXiv:0812.0376}.

\bibitem{Vallinotto:2010qm}
  A.~Vallinotto, S.~Dodelson and P.~Zhang,
  ``Magnification as a Tool in Weak Lensing,''
  \url{arXiv:1009.5590}.

\bibitem{menard03}
 B. M\'enard {\em et al.},
 ``Improving the accuracy of cosmic magnification statistics'',
 Astron. Astrophys. {\bf403}, 817 (2003) [arXiv:astro-ph/0210112].


\bibitem{ehlerss}
J. Ehlers and P. Schneider, ``Self-consistent probabilities for gravitational lensing in inhomogeneous universes", Astron. Astrophys. {\bf 168}, 57 (1986).

\bibitem{linder88}
 E.V. Linder,
 ``Light propagation in generalized Friedmann universes'',
 Astron. Astrophys. {\bf 206}, 190 (1988).

\bibitem{tomita}
 K. Tomita,
 ``Angular Diameter Distances in Clumpy Friedmann Universes'',
 Prog. Theor. Phys. {\bf100}, 79 (1998)
 \url{astro-ph/9806047}.

\bibitem{Mortsell:2001es}
  E.~Mortsell,
  ``The Dyer--Roeder distance-redshift relation in inhomogeneous universes,''
  \url{arXiv:astro-ph/0109197}.

\bibitem{Bolejko:2010nh}
  K.~Bolejko,
  ``Weak lensing and the Dyer--Roeder approximation,''
  Mon. Not. Roy. Astron. Soc. {\bf412} 1937 (2011)
  \url{arXiv:1011.3876}.

\bibitem{Wang:2011sb}
  Y.~Wang, C.~-H.~Chuang, P.~Mukherjee,
  ``A Comparative Study of Dark Energy Constraints from Current Observational Data,''
  [arXiv:1109.3172].

\bibitem{Mattsson:2007tj}
  T.~Mattsson,
  ``Dark energy as a mirage,''
  Gen.\ Rel.\ Grav.\  {\bf 42}, 567-599 (2010)
  [arXiv:0711.4264].

\bibitem{February:2009pv}
S. February, J. Larena, M. Smith and C. Clarkson, ``Rendering dark energy void", Mon. Not. Roy. Astron. Soc. {\bf 405}, 2231 (2010) [arXiv:0909.1479].

\bibitem{Brouzakis:2006dj}
  N.~Brouzakis, N.~Tetradis and E.~Tzavara,
  ``The Effect of Large-Scale Inhomogeneities on the Luminosity Distance,''
  JCAP {\bf 0702}, 013 (2007)
  [arXiv:astro-ph/0612179];
  N.~Brouzakis, N.~Tetradis and E.~Tzavara,
  ``Light Propagation and Large-Scale Inhomogeneities,''
  JCAP {\bf 0804}, 008 (2008)
  [arXiv:astro-ph/0703586].

\bibitem{Biswas:2007gi}
  T.~Biswas and A.~Notari,
  ``Swiss-Cheese Inhomogeneous Cosmology and the Dark Energy Problem,''
  JCAP {\bf 0806}, 021 (2008)
  [arXiv:astro-ph/0702555].

\bibitem{Marra:2007pm}
  V.~Marra, E.~W.~Kolb, S.~Matarrese and A.~Riotto,
  ``On cosmological observables in a swiss-cheese universe,''
  Phys.\ Rev.\  D {\bf 76}, 123004 (2007)
  [arXiv:0708.3622].

\bibitem{Vanderveld:2008vi}
  R.~A.~Vanderveld, E.~E.~Flanagan and I.~Wasserman,
  ``Luminosity distance in 'Swiss cheese' cosmology with randomized voids: I.
  Single void size,''
  Phys.\ Rev.\  D {\bf 78}, 083511 (2008)
  [arXiv:0808.1080].

\bibitem{Clifton:2009nv}
  T.~Clifton and J.~Zuntz,
  ``Hubble Diagram Dispersion From Large-Scale Structure,''
  [arXiv:0902.0726].

\bibitem{Valkenburg:2009iw}
  W.~Valkenburg,
  ``Swiss Cheese and a Cheesy CMB,''
  JCAP {\bf 0906}, 010 (2009)
  [arXiv:0902.4698].

\bibitem{Kantowski:2009jt}
  R.~Kantowski, B.~Chen and X.~Dai,
  ``Gravitational Lensing Corrections in Flat $\Lambda$CDM Cosmology,''
  Astrophys.\ J.\  {\bf 718}, 913 (2010)
  [arXiv:0909.3308].

\bibitem{Bolejko:2010eb}
  K.~Bolejko, M.~-N.~Celerier,
  ``Szekeres Swiss-Cheese model and supernova observations,''
  Phys.\ Rev.\  {\bf D82}, 103510 (2010)
  [arXiv:1005.2584].

\bibitem{Szybka:2010ky}
  S.~J.~Szybka,
  ``On light propagation in Swiss-Cheese cosmologies,''
  Phys.\ Rev.\  {\bf D84}, 044011 (2011)
  [arXiv:1012.5239].



\bibitem{Takahashi:2011qd}
  R.~Takahashi, M.~Oguri, M.~Sato and T.~Hamana,
  ``Probability Distribution Functions of Cosmological Lensing: Convergence,
  Shear, and Magnification,''
  [arXiv:1106.3823];
  S.~Hilbert, J.~R.~Gair, L.~J.~King,
  ``Reducing distance errors for standard candles and standard sirens with weak-lensing shear and flexion maps,''
 Mon. Not. Roy. Astron. Soc. {\bf 412}, 1023 (2011)
  [arXiv:1007.2468];
  S.~Hilbert, S.~D.~M.~White, J.~Hartlap, P.~Schneider,
  ``Strong lensing optical depths in a $\Lambda$CDM universe,''
  Mon. Not. Roy. Astron. Soc. {\bf 382}, 121 (2007)
  [arXiv:astro-ph/0703803].


\bibitem{Meures:2011gp}
  N.~Meures and M.~Bruni,
  ``Redshift and distances in a $\Lambda$CDM cosmology with nonlinear
  inhomogeneities,''
  [arXiv:1107.4433].

\bibitem{Krasinski:2010rc}
  A.~Krasinski, K.~Bolejko,
  ``Redshift propagation equations in the $\beta' \neq 0$ Szekeres models,''
  Phys.\ Rev.\  {\bf D83}, 083503 (2011)
  [arXiv:1007.2083].

\bibitem{Clifton:2009jw}
  T.~Clifton and P.~G.~Ferreira,
  ``Archipelagian Cosmology: Dynamics and Observables in a Universe with
  Discretized Matter Content,''
  Phys.\ Rev.\  D {\bf 80}, 103503 (2009)
  [arXiv:0907.4109];
  T.~Clifton, P.~G.~Ferreira,
  ``Errors in Estimating $\Omega_{\Lambda}$ due to the Fluid Approximation,''
  JCAP {\bf 0910}, 026 (2009) 
  [arXiv:0908.4488].
  
\bibitem{Bassett:2003vu}
  B.~A.~Bassett, M.~Kunz,
  ``Cosmic distance-duality as a probe of exotic physics and acceleration,''
  Phys.\ Rev.\  {\bf D69}, 101305 (2004).
  [astro-ph/0312443].
  
  \bibitem[Clarkson(2012)]{2012arXiv1204.5505C} C. Clarkson, 
Establishing homogeneity of the universe in the shadow of dark energy.\ 
[arXiv:1204.5505]. 

\bibitem[Busti and Lima(2012)]{2012arXiv1204.1083B}  V.~C. Busti and J.~A.~S. Lima,  Cosmological Tests Plagued by Small-Scale 
Inhomogeneities. [arXiv:1204.1083]. 


\end{references}
\end{document}